\documentclass[authoryear,12pt,onecolumn,3p]{elsarticle}%
\usepackage{amssymb}
\usepackage{amsmath}
\usepackage{amsfonts}
\usepackage{amsbsy}
\usepackage{graphicx}
\usepackage{pifont}
\usepackage[authoryear]{natbib}
\usepackage{geometry}
\usepackage{hyperref}
\usepackage{version}
\usepackage{setspace}%
\providecommand{\U}[1]{\protect\rule{.1in}{.1in}}
\journal{arXiv}
\newtheorem{theorem}{Theorem}
\newtheorem{proposition}[theorem]{Proposition}
\newtheorem{corollary}[theorem]{Corollary}

\newtheorem{remark}[theorem]{Remark}

\newtheorem{lemma}[theorem]{Lemma}
\newtheorem{definition}[theorem]{Definition}
\newtheorem{example}[theorem]{Example}
\excludeversion{comment2}
\excludeversion{comment3}
\begin{document}
\begin{frontmatter}
\title{Random partitions, potential, value, and externalities}
\author[hhl]{Andr{\'{e}} Casajus}
\ead{mail@casajus.de}
\ead[url]{www.casajus.de}
\author[was]{Yukihiko Funaki}
\ead{funaki@waseda.jp}
\ead[url]{yfunaki.blogspot.com}
\author[skk]{Frank Huettner\corref{cor}}
\ead{mail@frankhuettner.de}
\ead[url]{www.frankhuettner.de}
\cortext[cor]{Corresponding author}
\address[hhl]{\sl HHL Leipzig Graduate School of Management, Jahnallee~59, 04109~Leipzig, Germany}
\address[was]{\sl School of Political Science and Economics, Waseda University, Nishi-Waseda, Shinjuku-Ku, Tokyo, Japan}
\address[skk]{\sl SKK GSB, Sungkyunkwan University, 25-2, Sungkyunkwan-Ro, Jongno-gu, 110-745 Seoul, South Korea}
\begin{abstract}
The Shapley value equals a player's contribution to the potential of a game. The potential is a most natural one-number summary of a game, which can be computed as the expected accumulated worth of a random partition of the players. This computation integrates the coalition formation of all players and readily extends to games with externalities. We investigate those potential functions for games with externalities that can be computed this way. It turns out that the potential that corresponds to the MPW solution introduced by Macho-Stadler et al. (2007, J. Econ. Theory 135, 339--356) is unique in the following sense. It is obtained as the expected accumulated worth of a random partition, it generalizes the potential for games without externalities, and it induces a solution that satisfies the null player property even in the presence of externalities.
\end{abstract}
\end{frontmatter}

\newpage

\section{Introduction}

A significant part of the knowledge regarding the Shapley value is based on
understanding how it behaves in relation to subgames. The reference to
subgames might appear rather subtle as in the carrier axiom by
\citet{shapley1953}%
, or clearly emphasized as in other major characterizations, e.g., by
\cite{sobolev1975}%
,
\citet{myerson1980}%
, and
\citet{HarMas1989}%
. The latter concretize the idea that a player's payoff can be obtained as a
contribution to the \emph{game}, i.e., to a one-number summary of a game. They
call this one-number summary a \emph{potential}\textit{\emph{\/}} and obtain
uniqueness of both the potential as well as the Shapley value being a player's
contribution to the potential.
\citet{HarMas1989}
further emphasize the importance of the fact that the potential can be
computed in an intuitive way as the expected normalized worth in a game.
Whereas this computation considers coalitions in isolation,
\citet{casajus-potential}
demonstrates that the potential can alternatively be computed as the expected
accumulated worth of a random partition of \emph{all} players. This
perspective, which evaluates a coalition's worth in the context of the
coalition formation of the other players, is particularly striking if we want
to extend these ideas to a setup that includes externalities.

Cooperative games with e\emph{x}\textit{\emph{\/}}ternalities (henceforth
\emph{TUX games}, also known as games in partition function form) emerge as
more general approach to cooperative games with transferable utility
(henceforth \emph{TU games}\textit{\emph{\/}}). The presence of externalities
means that the worth generated by a coalition depends on the coalitions formed
by the other players
\citep{ThrLuc1963}%
. In search for a generalization of the Shapley value
\citep{shapley1953}%
, several solutions concepts for TUX games have been suggested. We highlight
\citet{DuEhKa2010}%
, who study subgames and the potential approach for TUX games and address the
following challenge. In the presence of externalities, one cannot simply read
off a subgame from the original game as it is the case for TU games.
\citet{DuEhKa2010}
argue that there are many ways to create subgames of a TUX game. This
motivates them to introduce the notion of restriction operators. A restriction
operator specifies how subgames are derived from TUX games. A restriction
operator is path independent if it does not matter in which order the players
are removed from a TUX game.
\citet{DuEhKa2010}
show that once a path independent restriction operator has been selected, then
there exists a unique potential for TUX games and a unique corresponding
solution for TUX\ games that gives a player the contribution to this potential.

In this paper, we connect these insights by asking wether there is a
one-number summary of each TUX game, such that (i) it is obtained as the
expected accumulated worth of some random partition of the players as in
\citet{casajus-potential}%
; (ii) it is induced by a plausible restriction operator, which connects TUX
games with their subgames as in
\citet{DuEhKa2010}%
; and (iii) a player's contribution to this number provides a generalization
of the Shapley value to TUX games, which inherits the basic properties of the
Shapley value
\citep{shapley1953}%
. The answer is affirmative and yields a new foundation of the MPW solution
introduced by
\citet{MSPCWe2007}%
.\footnote{This solution is also known as stochastic Shapley value
\citep{SkMiWo2018}%
\ or, somewhat imprecisely, as \emph{the}\textit{\emph{\/}} average Shapley
value.}

The remainder of this paper is structured as follows. In Section 2, we provide
the basic notation and revisit the relevant results. First, we focus on
TU\ games and the Shapley value, which is the contribution to the potential of
a TU\ game, which in turn can be obtained as the expected accumulated worth of
a random partition. Then, we provide the basic notation and revisit the
relevant results on TUX\ games, in particular the approach to subgames via
restriction operators by
\citet{DuEhKa2010}%
.

In Section 3, we provide our results. First, we revisit the result of
\citet{casajus-potential}%
, who has demonstrated that the Ewens distribution with mutation rate
$\theta=1$
\citep{ewens1972,KarMcG1972}
generates the potential for TU\ games as the expected accumulated worth of a
random partition of the player set. We demonstrate that there are further such
random partitions and investigate their properties.

We then obtain our first main result. Given a positive random partition that
generates the potential for TU\ games, we can construct a restriction operator
such that the expected accumulated worth of the random partition corresponds
to the potential for TUX games. Moreover, our construction is unique under
some mild assumptions on the restriction operator.

Our second main result comes from the study of those restriction operators
that we constructed and their induced solutions for TUX\ games. We find that
among these solutions, only the MPW\ solution satisfies the null player
property (or monotonicity) for TUX games. There, we also uncover the specific
restriction operator that naturally leads to the MPW\ solution. This
restriction operator is reminiscent of the Chinese restaurant process, which
is known to yield the Ewens distributions
(\citealp[11.19]{aldous1985}; \citealp[Equation~3.3]{pitman2006})%
.

We then discuss the following points. We contrast our approach and other
solutions for TUX games proposed in the literature and we discuss the
relationship between the condition that a random partition generates the
potential for TU games and the well-known conditional independence property.
Moreover, we discuss the fact that the restriction operator that yields the
MPW\ solution exhibits a surprising property, namely that it may convert a
null player into a non-null player when another player is removed. Yet, a
previously null player still obtains zero according to the MPW solution in the subgame.

Finally, we conclude on how our work may help to provide the basis for future
research. The appendix contains all the proofs.

\section{Preliminaries}

We introduce notation and results that are needed to understand our results in
several parts.\ First, we stay within the world of TU~games. Then, we include
externalities. The appendix includes additional notation that is needed for
the proofs.

\subsection{The Shapley value is a player's contribution to the expected
accumulated worth of a partition}

Let $\mathbf{U}$ be a finite set of players, the \textbf{universe of players}.
Throughout the paper, the cardinalities of coalitions $N,S,T,B,C\subseteq
\mathbf{U}$ are denoted by $n,$ $s,$ $t,$ $b,$ and $c,$ respectively. A
\textbf{partition} of $N\subseteq\mathbf{U}$ is a collection of non-empty
subsets of $N$ such that any two of them are disjoint and such their union is
$N.$ The \textbf{set of partitions} of $N$ is denoted by $\Pi\left(  N\right)
.$ Note that $\Pi\left(  \emptyset\right)  =\left\{  \emptyset\right\}  .$

A \textbf{random partition}\emph{ }for $\mathbf{U}$ is a family of probability
distributions $p=\left(  p_{N}\right)  _{N\subseteq\mathbf{U}}$ over
partitions, i.e., $p_{N}$ is a probability distribution over $\Pi\left(
N\right)  $. The random partition $p^{\star}$ given by
\begin{equation}
p_{N}^{\star}\left(  \pi\right)  :=\frac{\prod_{B\in\pi}\left(  b-1\right)
!}{n!}\text{\qquad for all }\pi\in\Pi\left(  N\right)  , \label{eq:p-star}%
\end{equation}
is known as the \textbf{Ewens distribution with mutation rate}\emph{ }%
$\theta=1$
\citep{ewens1972,KarMcG1972}%
. It plays a central role in the literature on random partitions
\citep{crane2016}%
.\footnote{\label{fn:ewens}The one-parametric family of Ewens distributions
$p^{\theta}$, $\theta>0$ is given by $p_{N}^{\theta}\left(  \pi\right)
:=\frac{\Gamma\left(  \theta\right)  \theta^{\left\vert \pi\right\vert }%
}{\Gamma\left(  \theta+n\right)  }\prod_{B\in\pi}\left(  b-1\right)  !$ for
$N\subseteq\mathbf{U},\ \pi\in\Pi\left(  N\right)  ,$ where $\Gamma$ denotes
the Gamma function. For all $\theta>0,$ $N\subseteq\mathbf{U},$ and $i,j\in
N,$ $i\neq j,$ we have $\sum_{\pi\in\Pi\left(  N\right)  :\pi\left(  i\right)
=\pi\left(  j\right)  }p_{N}^{\theta}\left(  \pi\right)  =\frac{1}{1+\theta}.$
Note that a mutation rate of~$1$ corresponds to a probability of $1/2$ of two
arbitrary players belonging to the same block of a partition.}

For $\pi\in\Pi\left(  N\setminus\left\{  i\right\}  \right)  ,$ \textbf{adding
the player} $i\in N$ to block $B\in\pi$ is denoted by $\pi_{+i\leadsto B}%
\in\Pi\left(  N\right)  ,$
\[
\pi_{+i\leadsto B}:=\left(  \pi\setminus\left\{  B\right\}  \right)
\cup\left\{  B\cup\left\{  i\right\}  \right\}  ;
\]
adding player~$i$ as a singleton is denoted by $\pi_{+i\leadsto\emptyset}%
\in\Pi\left(  N\right)  ,$ i.e., $\pi_{+i\leadsto\emptyset}:=\pi\cup\left\{
\left\{  i\right\}  \right\}  $.

A cooperative game with transferable utility, henceforth \textbf{TU game}
(also known as a game in characteristic function form), for a player set
$N\subseteq\mathbf{U}$ is given by its \textbf{characteristic function}
$v:2^{N}\rightarrow\mathbb{R}$, $v\left(  \emptyset\right)  =0,$ which assigns
a \textbf{worth} to each coalition $S\subseteq N$. Let $\mathbb{V}\left(
N\right)  \ $denote the set of all TU games for $N$ and let $\mathbb{V}$
denote the set of all TU games. For $v\in\mathbb{V}\left(  N\right)  $ and
$i\in N,$ the \textbf{subgame~}$v_{-i}$ on the player set $N\setminus\left\{
i\right\}  $ is simply the \textbf{restriction} of $v$ and given by
$v_{-i}\left(  S\right)  =v\left(  S\right)  $ for all $S\subseteq
N\setminus\left\{  i\right\}  .$ A one-point \textbf{solution for TU\ games}
is an operator $\varphi\mathbf{\ }$that assigns a payoff vector $\varphi
\left(  v\right)  \in\mathbb{R}^{N}$ to every TU game $v\in\mathbb{V}\left(
N\right)  $ and player set $N\subseteq\mathbf{U}.$

The Shapley value
\citep{shapley1953}%
\ probably is the most eminent one-point solution concept for TU games.
Besides its original axiomatic foundation by Shapley himself, alternative
foundations of different types have been suggested later on. Important direct
axiomatic characterizations are due to
\citet{myerson1980}%
\ and
\citet{young1985}%
. The \textbf{Shapley value}, $\mathrm{Sh}$,\ is given by%
\begin{equation}
\mathrm{Sh}_{i}\left(  v\right)  :=\sum_{S\subseteq N\setminus\left\{
i\right\}  }\frac{s!\left(  n-s-1\right)  !}{n!}\left(  v\left(  S\cup\left\{
i\right\}  \right)  -v\left(  S\right)  \right)  \label{eq:Sh}%
\end{equation}
for all $N\subseteq\mathbf{U},\ v\in\mathbb{V}\left(  N\right)  ,$ and $i\in
N.$%

\citet{HarMas1989}%
\ suggest an indirect characterization of the Shapley value by means of a
potential (function) for TU games. A \textbf{potential} is a mapping
$\mathrm{Pot}:\mathbb{V}\rightarrow\mathbb{R}$ that satisfies the following
two properties. Zero-normalization: For $v_{\emptyset}\in\mathbb{V}\left(
\emptyset\right)  ,$ we have $\mathrm{Pot}\left(  v_{\emptyset}\right)  =0,$
i.e., the unique game for the empty player set has a zero potential.$\ $%
Efficiency: For all $N\subseteq\mathbf{U},$ $v\in\mathbb{V}\left(  N\right)  $
$,$we have%
\begin{equation}
\sum_{i\in N}\left[  \mathrm{Pot}\left(  v\right)  -\mathrm{Pot}\left(
v_{-i}\right)  \right]  =v\left(  N\right)  . \label{eq:def-pot}%
\end{equation}
It turns out that there exists a unique potential and that a player's
contribution to the potential equals this player's Shapley payoff.\footnote{%
\citet{CalSan1997}%
\ and
\citet{ortmann1998}%
\ generalize the notion of a potential to non-efficient solutions.}

\begin{theorem}
[%
\citealp{HarMas1989}%
]\label{thm:HM1989}There exists a unique potential for TU games and we have%
\begin{equation}
\mathrm{Sh}_{i}\left(  v\right)  =\mathrm{Pot}\left(  v\right)  -\mathrm{Pot}%
\left(  v_{-i}\right)  \qquad\text{for all }N\subseteq\mathbf{U}%
,\ v\in\mathbb{V}\left(  N\right)  ,\ \text{and }i\in N. \label{eq:HMSh=pot}%
\end{equation}

\end{theorem}

%

\citet{casajus-potential}%
\ establishes that the potential can be calculated as the expected accumulated
worth of a random partition.

\begin{theorem}
[%
\citealp{casajus-potential}%
]\label{thm:Pot-p-star}For all $N\subseteq\mathbf{U},\ v\in\mathbb{V}\left(
N\right)  $, we have%
\begin{equation}
\mathrm{Pot}\left(  v\right)  =\sum_{\pi\in\Pi\left(  N\right)  }p_{N}^{\star
}\left(  \pi\right)  \sum_{B\in\pi}v\left(  B\right)  , \label{eq:Pot-p-star}%
\end{equation}
where the random partition $p^{\star}$ is the Ewens distribution with mutation
rate\emph{ }$\theta=1$ given in (\ref{eq:p-star}).
\end{theorem}

As a consequence, we can understand the Shapley value as a player's
contribution to the expected accumulated worth of a partition. The Shapley
value, $\mathrm{Sh}$,\ is given by%
\begin{equation}
\mathrm{Sh}_{i}\left(  v\right)  =\sum_{\pi\in\Pi\left(  N\right)  }%
p_{N}^{\star}\left(  \pi\right)  \sum_{B\in\pi}v\left(  B\right)  -\sum
_{\tau\in\Pi\left(  N\setminus\left\{  i\right\}  \right)  }p_{N\setminus
\left\{  i\right\}  }^{\star}\left(  \tau\right)  \sum_{C\in\tau}v_{-i}\left(
C\right)  \label{eq:Sh=EVpartition}%
\end{equation}
for all $N\subseteq\mathbf{U},$ $v\in\mathbb{V}\left(  N\right)  ,$ and $i\in
N.$ Note that the right-hand-sides of (\ref{eq:HMSh=pot}) and
(\ref{eq:Sh=EVpartition}) refer to games with varying player sets. In this
sense, the Shapley value emerges as a player's \emph{contribution to a game}
and not as the weighted contribution to coalitions. Operating on varying
player sets is also useful for implementations and bargaining foundations of
the value, which often rest on the idea that some player's productivity is
bought out by other players, stepwise reducing the player set (prominent
examples include
\citet{KriSer1995}%
,
\citet{HarMas1996}%
, {%
\citet{gul1989}%
}, {%
\citet{StoZwi1996}%
}, {%
\citet{PerWet2001}%
}, {%
\citet{McQSug2016}%
}, {%
\citet{BruGauMen2018}%
}).

Thinking of a game in terms of the expected accumulated worth of a partition
is particularly striking if we keep in mind that the value of a coalition
should be understood in context of the coalition structure of outsiders. This
is important in the context of externalities, i.e., if the value of a
coalition depends on the coalitions formed by the other players. Naturally, we
arrive at the question: \emph{Can (\ref{eq:Sh=EVpartition}) be generalized
towards a setup that takes externalities into account?}

Of course, intuitive formulas for the potential as in (\ref{eq:Pot-p-star})
are of interest by itself. For example,
\citet[Proposition~2.4]{HarMas1989}
emphasize that the per-capita potential of a game is the expected per-capita
worth of a standard random coalition. First, any coalition size is chosen with
the same probability. Second, any coalition of a given size is chosen with the
same probability. That is, we have%
\begin{equation}
\frac{\mathrm{Pot}\left(  v\right)  }{n}=\sum_{S\subseteq N:S\neq\emptyset
}\frac{1}{n}\frac{s!\left(  n-s\right)  !}{n!}\frac{v\left(  S\right)  }%
{s}\qquad\text{for all }N\subseteq\mathbf{U},~v\in\mathbb{V}\left(  N\right)
. \label{eq:Pot2}%
\end{equation}
They conclude that this formula indicates that \textquotedblleft... the
potential provides a most natural one-number summary of the
game\textquotedblright\
\citep[p.~592]{HarMas1989}%
. Likewise, a potential for games with externalities that exhibits some
intuitive formula shall be appreciated.

Finally, we shall notice that the random partition $p^{\star}$ given in
(\ref{eq:p-star}) is the consequence of a stochastic coalition formation
process known as the \textquotedblleft\emph{(uniform) Chinese restaurant
process}\textquotedblright\
(\citealp[11.19]{aldous1985}; \citealp[Equation~3.3]{pitman2006})%
. It can be described as follows. Consider a restaurant\ with $n$ round
tables, each seating up to $n$~persons. The players arrive at the restaurant
in any order with the same probability. The first player takes a seat at one
of the tables. Any following player takes seat at an empty table or joins any
of the already present players with the \emph{same}\textit{\emph{\/}}
probability. If, for example, there already are two non-empty tables, the
first one with two players and the second one with one player, then the fourth
player joins the first table with probability 2/4, the second table with
probability 1/4, and takes seat at an empty table with probability 1/4. This
motivates the use of \textquotedblleft\emph{uniform}\textit{\emph{\/}%
}\textquotedblright\ above. This process induces a random partition where each
table will be understood as a block of a partition. Concretely, after $n$
players have entered the room and taken a seat, each occupied table will be
understood as a block of $\pi\in\Pi\left(  N\right)  $ and $\pi$ has the
probability $p_{N}^{\star}\left(  \pi\right)  $. Hence, the probability
distributions $p_{N}^{\star}$ and $p_{N\setminus\left\{  i\right\}  }^{\star}$
are connected by%
\begin{equation}
p_{N}^{\star}\left(  \pi_{+i\leadsto\emptyset}\right)  =\frac{1}%
{n}p_{N\setminus\left\{  i\right\}  }^{\star}\left(  \pi\right)
\qquad\text{and}\qquad p_{N}^{\star}\left(  \pi_{+i\leadsto B}\right)
=\frac{b}{n}p_{N\setminus\left\{  i\right\}  }^{\star}\left(  \pi\right)  .
\label{eq:p-con}%
\end{equation}

Insertion of (\ref{eq:p-con}) into (\ref{eq:Sh=EVpartition}) yields a formula
of the Shapley value as the expected marginal contribution to coalitions over
all partitions. The Shapley value, $\mathrm{Sh}$,\ is given by%
\begin{equation}
\mathrm{Sh}_{i}\left(  v\right)  =\sum_{\pi\in\Pi\left(  N\setminus\left\{
i\right\}  \right)  }p_{N\setminus\left\{  i\right\}  }^{\star}\left(
\pi\right)  \left(  \frac{1}{n}\left(  v\left(  \left\{  i\right\}  \right)
-v\left(  \left\{  \emptyset\right\}  \right)  \right)  +\sum_{S\in\pi}%
\frac{s}{n}\left(  v\left(  S\cup\left\{  i\right\}  \right)  -v\left(
S\right)  \right)  \right)  \label{eq:Shchinese}%
\end{equation}
for all $N\subseteq\mathbf{U},$ $v\in\mathbb{V}\left(  N\right)  ,$ and $i\in
N.$ In words, the Shapley value is a player's expected marginal contribution
to a table when entering the Chinese restaurant process last. This might, for
instance, be helpful for computations of the Shapley value based on Monte
Carlo sampling from random partitions.\footnote{Standard Monte Carlo
algorithms to compute the Shapley value rely on sampling from rank orders,
e.g.,
\citet{CaGoTe2009}%
; however, exploiting particular formulas has shown to be more efficient for
some games, e.g.,
\citet{MiAaSzRaJe2013}
or
\citet{Lundbergetal2020}%
.}

In this paper, however, we primarily focus on formulas that investigate a
player's \emph{contribution to a game}\textit{\emph{\/}} as in
(\ref{eq:Sh=EVpartition}), rather than on formulas that exhibit a player's
\emph{contribution to coalition}\textit{\emph{\/}} as for instance in
(\ref{eq:Shchinese}). For the latter, we refer to
\citep{SkMiWo2018}%
\ and
\citet{SkiMic2020}%
, who study games with externalities and solutions thereof in the context of
stochastic coalition formation processes.

\subsection{Games with externalities and subgames\label{sec:tux}}

A TU game with externalities, henceforth \textbf{TUX game }(also known as
\textbf{game in partition function form}), for a player set $N\subseteq
\mathbf{U}$ is given by its \textbf{partition function} $w:\mathcal{E}\left(
N\right)  \rightarrow\mathbb{R},$ where $\mathcal{E}\left(  N\right)  $
denotes the set of \textbf{embedded coalitions} $\left(  S,\pi\right)  $ for
$N$ given by
\[
\mathcal{E}\left(  N\right)  :=\left\{  \left(  S,\pi\right)  \mid S\subseteq
N\text{ and }\pi\in\Pi\left(  N\setminus S\right)  \right\}  ,
\]
and with $w\left(  \emptyset,\pi\right)  =0\ $for$\ $all$\ \pi\in\Pi\left(
N\right)  $. We denote the set of all TUX games for a player set $N$ by
$\mathbb{W}\left(  N\right)  $ and the set of all TUX games by\ $\mathbb{W}$.

If there are no externalities, that is, if $w\left(  S,\pi\right)  =w\left(
S,\tau\right)  $ for all $S\subseteq N$ and all $\pi,\tau\in\Pi\left(
N\setminus S\right)  $, then the game $w$ effectively is just a TU game. In
this sense TU games are special TUX games, $\mathbb{V\subseteq W}$. An example
of a game without externalities is the \textbf{null game} $\mathbf{0}^{N}%
\in\mathbb{W}\left(  N\right)  ,$ $N\subseteq\mathbf{U}$, given by
$\mathbf{0}^{N}\left(  S,\pi\right)  =0$ for all $\left(  S,\pi\right)
\in\mathcal{E}\left(  N\right)  .$

A one-point \textbf{solution for TUX games} is an operator $\varphi
\mathbf{\ }$that assigns a payoff vector $\varphi\left(  w\right)
\in\mathbb{R}^{N}$ to every TUX game $w\in\mathbb{W}\left(  N\right)  $ and
player set $N\subseteq\mathbf{U}.$ Whereas there is no eminent solution for
TUX games, the literature provides numerous \textbf{generalizations of the
Shapley value}, i.e., solutions for TUX games that coincide with the Shapley
value on TU\ games, $\varphi\left(  v\right)  =\mathrm{Sh}\left(  v\right)  $
for all $v\in\mathbb{V}\left(  N\right)  \subseteq\mathbb{W}\left(  N\right)
,\ N\subseteq\mathbf{U}.$ Examples include
\citet{myerson1977pffg}%
,%
\citet{bolger1989}%
,
\citet{AlArRu2005}%
,
\citet{PhaNor2007}%
,
\citet{mcquillin2009}%
, and
\citet{DuEhKa2010}%
; see
\citet{koszy2018}%
\ for a survey. We highlight the solution introduced by
\citet{MSPCWe2007}%
, which is obtained as the Shapley value of some average TU\ game, i.e.,%
\begin{equation}
\mathrm{MPW}\left(  w\right)  :=\mathrm{Sh}\left(  \bar{v}_{w}^{\star}\right)
\qquad\text{for all }N\subseteq\mathbf{U},~w\in\mathbb{W}\left(  N\right)  ,
\label{eq:Sh-star}%
\end{equation}
where the average TU\ game $\bar{v}_{w}^{\star}\in\mathbb{V}\left(  N\right)
$, in which each coalition~$S$ gets the expected value of $w\left(
S,\pi\right)  $ over all partitions $\pi\in\Pi\left(  N\setminus S\right)  $
using the random partition $p^{\star}$, is given by
\begin{equation}
\bar{v}_{w}^{\star}\left(  S\right)  :=\sum_{\pi\in\Pi\left(  N\setminus
S\right)  }p_{N\setminus S}^{\star}\left(  \pi\right)  w\left(  S,\pi\right)
\qquad\text{for all }S\subseteq N. \label{eq:averagegame}%
\end{equation}

Both the potential and the understanding of the Shapley value as a player's
contribution to the expected accumulated worth of a partition rely on the
notion of subgames, which result from reducing the player set. For TU games,
there is an obvious way to define subgames. In contrast, the notion of a
subgame is less obvious for TUX games, since we cannot simply read it off the
original game. When player $i$ is removed from the TUX game $w$, we have to
specify the worth of each embedded coalition \textquotedblleft$w_{-i}\left(
S,\pi\right)  $\textquotedblright\ in the TUX game $w_{-i}$ without player
$i$. Other than for TU games, it is unclear how \textquotedblleft%
$w_{-i}\left(  S,\pi\right)  $\textquotedblright\ is obtained from $w$.%

\citet{DuEhKa2010}
introduce the concept of a \textbf{restriction operator} $r$ that specifies
for each TUX game $w\in\mathbb{W}\left(  N\right)  ,$ $N\subseteq\mathbf{U},$
and player $i\in N$ a \textquotedblleft subgame\textquotedblright\ $w_{-i}%
^{r}\in\mathbb{W}\left(  N\setminus\left\{  i\right\}  \right)  $, where the
worth $w_{-i}^{r}\left(  S,\pi\right)  $ in the subgame only depends on the
worths of those embedded coalitions in the original game that are obtained
when player $i$ is added to a block in $\pi$ or stays alone. That is, a
restriction operator satisfies the following property.

\smallskip

\noindent\textbf{Restriction, RES.}$\;$For all $N\subseteq\mathbf{U},$
$w,w^{\prime}\in\mathbb{W}\left(  N\right)  $, $i\in N,$ and $\left(
S,\pi\right)  \in\mathcal{E}\left(  N\setminus\left\{  i\right\}  \right)  $
such that $w\left(  S,\pi_{+i\leadsto B}\right)  =w^{\prime}\left(
S,\pi_{+i\leadsto B}\right)  $ for all $B\in\pi\cup\left\{  \emptyset\right\}
,$ we have $w_{-i}^{r}\left(  S,\pi\right)  =\left(  w^{\prime}\right)
_{-i}^{r}\left(  S,\pi\right)  .$

\smallskip

Another crucial property of restriction operators is called path independence.
A restriction operator is path independent if the order in which the players
are removed is irrelevant.

\smallskip

\noindent\textbf{Path independence, PI.}$\;$For all $N\subseteq\mathbf{U},$
$w\in\mathbb{W}\left(  N\right)  ,$ and $i,j\in N,$ $i\neq j,$ we have
$(w_{-i}^{r})_{-j}^{r}=(w_{-j}^{r})_{-i}^{r}.$

\smallskip

\noindent Note that we can refer to subgames obtained by the elimination of
multiple players whenever $r$ is path independent. That is, the subgames
$w_{-T}^{r},$ $T\subseteq N,$ are well-defined for path independent
restriction operators.

With the ability to remove players from a game,
\citet{DuEhKa2010}
can\ generalize the notion of potential functions to TUX games$.$ Given a path
independent restriction operator $r$, define the $r$\textbf{-potential} by
replacing $v_{-i}$ in (\ref{eq:def-pot}) with $w_{-i}^{r}$. That is, an
$r$-potential is a mapping $\mathrm{Pot}^{r}:\mathbb{W}\rightarrow\mathbb{R}$
that satisfies the following two properties: Zero-normalization: For
$\mathbf{0}^{\emptyset}\in\mathbb{W}\left(  \emptyset\right)  ,$ we have
$\mathrm{Pot}^{r}\left(  \mathbf{0}^{\emptyset}\right)  =0,$ i.e., the unique
game for the empty player set has a zero potential. Efficiency: For all
$N\subseteq\mathbf{U}$, $w\in\mathbb{W}\left(  N\right)  ,$ we have%
\begin{equation}
\sum_{i\in N}\left[  \mathrm{Pot}^{r}\left(  w\right)  -\mathrm{Pot}%
^{r}\left(  w_{-i}^{r}\right)  \right]  =w\left(  N,\emptyset\right)  \text{.}
\label{eq:r-pot2}%
\end{equation}
This recursively determines a well-defined and unique potential for each path
independent restriction operator $r$.

Similarly to the Shapley value, which is the contribution of a player to the
potential of a TU game, each path independent restriction operator induces a
unique solution for TUX games, the $r$\textbf{-Shapley value}, which is the
contribution of a player to the $r$-potential of a TUX game. The connection
between the $r$-Shapley value and the Shapley value is via auxiliary TU games
$v_{w}^{r}\in\mathbb{V}\left(  N\right)  $:

\begin{enumerate}
\item For a given TUX game $w\in\mathbb{W}\left(  N\right)  ,$ $N\subseteq
\mathbf{U}$, we first compute the auxiliary TU\ game $v_{w}^{r}\in
\mathbb{V}\left(  N\right)  $, in which each coalition $S$ generates the worth
generated by the grand coalition if we remove all the other players
$N\setminus S$ from the game. That is, the auxiliary TU\ game is defined by%
\begin{equation}
v_{w}^{r}\left(  S\right)  :=w_{-N\setminus S}^{r}\left(  S,\emptyset\right)
\qquad\text{for all }S\subseteq N. \label{eq:w-tilde-r}%
\end{equation}

\item Second, we apply the Shapley value to this auxiliary TU game in order to
obtain the $r$-Shapley value,
\begin{equation}
\mathrm{Sh}^{r}\left(  w\right)  :=\mathrm{Sh}\left(  v_{w}^{r}\right)
\qquad\text{for all }w\in\mathbb{W}\left(  N\right)  ,~N\subseteq\mathbf{U}.
\label{eq:r-Shapleyvalue}%
\end{equation}

\end{enumerate}

%

\citet{DuEhKa2010}%
\ find that the $r$-potential of the TUX game $w$ is just the potential of the
auxiliary TU game $v_{w}^{r}$ and that the $r$-Shapley value is a player's
contribution to the $r$-potential of a TUX game.

\begin{theorem}
[Dutta et al., 2010]\label{thm:exist-PF-pot}For every path independent
restriction operator $r$, there exists a unique $r$-potential $\mathrm{Pot}%
^{r}$. It is given by
\begin{equation}
\mathrm{Pot}^{r}\left(  w\right)  =\mathrm{Pot}\left(  v_{w}^{r}\right)
\qquad\text{for all }N\subseteq\mathbf{U},~w\in\mathbb{W}\left(  N\right)  ,
\label{eq:Pot-r-w}%
\end{equation}
where $v_{w}^{r}$ is defined in (\ref{eq:w-tilde-r}). Moreover, we have%
\begin{equation}
\mathrm{Sh}_{i}^{r}\left(  w\right)  =\mathrm{Pot}^{r}\left(  w\right)
-\mathrm{Pot}^{r}\left(  w_{-i}^{r}\right)  \qquad\text{for all }%
N\subseteq\mathbf{U},~w\in\mathbb{W}\left(  N\right)  ,\text{ and }i\in N.
\label{eq:Sh(w-tilde-r)}%
\end{equation}

\end{theorem}

Equipped with these insights, we can now proceed to our research questions.

\section{Results}

In this section, we investigate if it is possible to summarize each TUX game
by a single number such that this number is obtained as the $r$-potential of a
TUX\ game for some restriction operator $r$, and such that this number is also
obtained as the \textbf{expected accumulated worth of some random
partition}~$p$, given by%
\begin{equation}
\mathrm{E}_{p}\left(  w\right)  :=\sum_{\pi\in\Pi\left(  N\right)  }%
p_{N}\left(  \pi\right)  \sum_{S\in\pi}w\left(  S,\pi\setminus\left\{
S\right\}  \right)  \text{\qquad for all }N\subseteq\mathbf{U},~w\in
\mathbb{W}\left(  N\right)  \text{.} \label{eq:def-Ep}%
\end{equation}
Moreover, we want this one-number summary to equal the potential for TU~games
if there are no externalities. Hence, we are interested in random partitions
that satisfy the following property.

\smallskip

\noindent\textbf{Potential generating, GEN.}$\;$For all TU~games
$v\in\mathbb{V}\left(  N\right)  \subseteq\mathbb{W}\left(  N\right)  $ and
player sets $N\subseteq\mathbf{U}$, we have $\mathrm{E}_{p}\left(  v\right)
=\mathrm{Pot}\left(  v\right)  $, where the potential for TU\ games
$\mathrm{Pot}\left(  v\right)  $ is defined in~(\ref{eq:def-pot}).

\smallskip

\noindent The answer is affirmative and will be constructive, i.e., we can
provide a formula for the respective restriction operators. This leads to our
final question: Do the induced $r$-Shapley values preserve standard properties
of the Shapley value. As it turns out, this is only the case for the
MPW~solution introduced by
\citet{MSPCWe2007}%
.

We structure our journey as follows. To begin with, we shed some light on
random partitions that generate the potential and motivate further plausible
conditions on random partitions. Thereafter, we describe the structure of
restriction operators $r^{p}$, such that $\mathrm{Pot}^{r^{p}}=\mathrm{E}_{p}$
for some random partition $p$ that generates the potential for TU~games
(\textbf{GEN}). Finally, we identify the MPW solution as the only resulting
$r^{p}$-Shapley value that satisfies the null player property or monotonicity
in presence of externalities.

\subsection{Random partitions that generate the potential for
TU~games\label{sec:GEN}}

For games with more than four players, the Ewens distribution $p^{\star}$ is
not the only random partition that generates the potential for TU~games
(\textbf{GEN}). More precisely, the following proposition describes the
characteristics of potential-generating random partitions.

\begin{proposition}
\label{pro:p=/p-star}The following statements are equivalent:

(i) The random partition $p$ generates the potential for TU\ games
(\textbf{GEN}).

(ii) The random partition $p$ satisfies%
\begin{equation}
\sum_{\pi\in\Pi\left(  N\right)  :T\in\pi}p_{N}\left(  \pi\right)
=\dfrac{\left(  n-t\right)  !\left(  t-1\right)  !}{n!}\text{\qquad for
all}\ N\subseteq\mathbf{U},\text{ }T\subseteq N,\text{ }T\neq\emptyset.
\label{eq:prop-p-gives-Pot-for-Dirac}%
\end{equation}

(iii) The random partition $p$ satisfies%
\begin{equation}
\sum_{\pi\in\Pi\left(  \left(  N\setminus\left\{  i\right\}  \right)
\setminus S\right)  }p_{N\setminus\left\{  i\right\}  }\left(  \left\{
S\right\}  \cup\pi\right)  =\frac{n}{n-s}\sum_{\pi\in\Pi\left(  \left(
N\setminus\left\{  i\right\}  \right)  \setminus S\right)  }\sum_{B\in\pi
\cup\left\{  \emptyset\right\}  }p_{N}\left(  \left\{  S\right\}  \cup
\pi_{+i\leadsto B}\right)  \label{eq:p-reduction}%
\end{equation}
for all $N\subseteq\mathbf{U},\ i\in N,$ and $S\subseteq N\setminus\left\{
i\right\}  ,$ $S\neq\emptyset.$
\end{proposition}

Whereas condition (ii) gives an explicit expression for the probability of a
particular coalition to appear in a randomly chosen partition, condition (iii)
offers a recursive relation. The latter will be useful for the construction of
a restriction operator that \textquotedblleft behaves well\textquotedblright%
\ on TU~games. Note that these conditions already determine the probabilities
$p_{N}\left(  \left\{  N\right\}  \right)  $ and $p_{N}\left(  \left\{
N\setminus\left\{  i\right\}  ,\left\{  i\right\}  \right\}  \right)  $.
Moreover, for games with up to three players, the random partition $p$ needs
to coincide with the Ewens distribution~$p^{\star}$.

\begin{corollary}
\label{cor:consequence-of-gen}For any random partition $p$ that generates the
potential for TU\ games (\textbf{GEN}) the following holds true:

(i) we have $p_{N}\left(  \left\{  N\right\}  \right)  =\frac{1}{n}$ and
$p_{N}\left(  \left\{  N\setminus\left\{  i\right\}  ,\left\{  i\right\}
\right\}  \right)  =\frac{1}{n\left(  n-1\right)  }$ for all $N\subseteq
\mathbf{U}$ and $i\in N.$

(ii) For all $N\subseteq\mathbf{U}$ such that $n\leq3$, we have $p_{N}\left(
\pi\right)  =p_{N}^{\star}\left(  \pi\right)  $ for all $\pi\in\Pi\left(
N\right)  $, where $p^{\star}$ is given in (\ref{eq:p-star}).
\end{corollary}

In games with more than three players, we obtain additional degrees of
freedom. For instance, every random partition $p^{c}$, $-\frac{1}{24}\leq
c\leq\frac{1}{8}$, that assigns the following probabilities to partitions of
four players, $\mathbf{U}=\left\{  i,j,k,\ell\right\}  $, generates the
potential for TU\ games:%
\begin{equation}
p_{4}^{c}\left(  \pi\right)  =\left\{
\begin{array}
[c]{ll}%
p_{4}^{\star}\left(  \pi\right)  +\frac{1}{3}c, & \pi=\left\{  \left\{
i,j\right\}  ,\left\{  k,\ell\right\}  \right\}  ,\\
p_{4}^{\star}\left(  \pi\right)  -\frac{1}{3}c, & \pi=\left\{  \left\{
i,j\right\}  ,\left\{  k\right\}  ,\left\{  \ell\right\}  \right\}  ,\\
p_{4}^{\star}\left(  \pi\right)  +c, & \pi=\left\{  \left\{  i\right\}
,\left\{  j\right\}  ,\left\{  k\right\}  ,\left\{  \ell\right\}  \right\}
,\\
p_{4}^{\star}\left(  \pi\right)  , & \text{otherwise.}%
\end{array}
\right.  \label{eq:pot4players}%
\end{equation}
For details on how to extend (\ref{eq:pot4players}) to larger games, we refer
to the appendix.

Note that setting $c=\frac{1}{8}$ and $c=-\frac{1}{24}$ in
(\ref{eq:pot4players}) gives random partitions that assign zero probability to
partitions of types $\left\{  \left\{  i,j\right\}  ,\left\{  k\right\}
,\left\{  \ell\right\}  \right\}  $ and $\left\{  \left\{  i\right\}
,\left\{  j\right\}  ,\left\{  k\right\}  ,\left\{  \ell\right\}  \right\}  $,
respectively. In what follows, we might not be interested in such random
partitions. Instead, we usually want that the expected accumulated worth of a
random partition makes use of all the information of a game. Hence, we may
assume the following positivity condition on random partitions.

\smallskip

\noindent\textbf{Positivity, POS.}$\;$For all $N\subseteq\mathbf{U}$ and
$\pi\in\Pi\left(  N\right)  ,$ we have $p_{N}\left(  \pi\right)  >0.$

\smallskip

\noindent We discuss the role of positivity in our paper in detail after our
final result (Theorem~\ref{thm:MPW}) at the end of Subsection~\ref{sec:MPW}.

\subsection{Potentials for TUX games equaling the expected accumulated worth
of a partition\label{sec:r-p}}

We are ready to develop our first main result. It describes the structure of
those restriction operators that induce a potential that both generalizes the
potential for TU\ games and is given as the expected accumulated worth of a
random partition. Suppose we are given some positive random partition~$p$ that
generates the potential for TU~games. In search of a suitable restriction
operator $r^{p}$, we need to preserve the natural restriction on TU~games,
i.e., $v_{-i}^{r^{p}}\left(  S,\pi\right)  =v\left(  S,\pi_{+i\leadsto
B}\right)  $ for all $v\in\mathbb{V}\left(  N\right)  \subseteq\mathbb{W}%
\left(  N\right)  $, $\left(  S,\pi\right)  \in\mathcal{E}\left(
N\setminus\left\{  i\right\}  \right)  $, and $B\in\pi\cup\left\{
\emptyset\right\}  $. By Proposition~\ref{pro:p=/p-star}~(iii), we then have%
\begin{align*}
&  \sum_{\tau\in\Pi\left(  \left(  N\setminus\left\{  i\right\}  \right)
\setminus S\right)  }p_{N\setminus\left\{  i\right\}  }\left(  \left\{
S\right\}  \cup\pi\right)  v_{-i}^{r^{p}}\left(  S,\pi\right) \\
&  \left.  \qquad\qquad\qquad=\right.  \frac{n}{n-s}\sum_{\tau\in\Pi\left(
\left(  N\setminus\left\{  i\right\}  \right)  \setminus S\right)  }\sum
_{B\in\pi\cup\left\{  \emptyset\right\}  }p_{N}\left(  \left\{  S\right\}
\cup\pi_{+i\leadsto B}\right)  v\left(  S,\pi_{+i\leadsto B}\right)  .
\end{align*}
This indicates that the following restriction operator works well for TU~games
$\mathbb{V}\left(  N\right)  \subseteq\mathbb{W}\left(  N\right)  $.

\begin{definition}
\label{def:rp}Let~$p$ be a positive (\textbf{POS}) random partition that
generates the potential for TU~games (\textbf{GEN}). We define the restriction
operator~$r^{p}$ by%
\begin{equation}
w_{-i}^{r^{p}}\left(  S,\pi\right)  :=\frac{n}{n-s}\sum_{B\in\pi\cup\left\{
\emptyset\right\}  }\frac{p_{N}\left(  \left\{  S\right\}  \cup\pi_{+i\leadsto
B}\right)  }{p_{N\setminus\left\{  i\right\}  }\left(  \left\{  S\right\}
\cup\pi\right)  }w\left(  S,\pi_{+i\leadsto B}\right)  \label{eq:r-p}%
\end{equation}
for all $N\subseteq\mathbf{U},$ $w\in\mathbb{W}\left(  N\right)  ,$ $i\in N,$
and $\left(  S,\pi\right)  \in\mathcal{E}\left(  N\setminus\left\{  i\right\}
\right)  $.
\end{definition}

It remains to be clarified whether this restriction operator also has
plausible properties for games with externalities. This is the content of the
next theorem, which further emphasizes the \emph{uniqueness}\textit{\emph{\/}}
of this restriction operator. We prepare this result we a condition on
restriction operators, which ensures that null games remain null games after
the removal of a player.

\smallskip

\noindent\textbf{Preservation of null games, PNG.}$\;$For all $N\subseteq
\mathbf{U}$ and $i\in N,$ we have $\left(  \mathbf{0}^{N}\right)  _{-i}%
^{r}=\mathbf{0}^{N\setminus\left\{  i\right\}  }.$

\smallskip

\noindent Ruling out oddly creative restriction operators with this condition
now leads us to the restriction operator~$r^{p}$.

\begin{theorem}
\label{thm:canext}Let $p$ be a positive random partition (\textbf{POS}) that
generates the potential for TU games (\textbf{GEN}). A restriction operator
$r$ is path independent (\textbf{PI}), preserves null games (\textbf{PNG}),
and satisfies $\mathrm{Pot}^{r}=\mathrm{E}_{p}$, i.e., its induced
$r$-potential defined in (\ref{eq:Pot-r-w}) coincides with the expected
accumulated worth of the random partition $p$ given in (\ref{eq:def-Ep}), if
and only if $r=r^{p}$.
\end{theorem}

Theorem~\ref{thm:canext} is useful for multiple reasons. It clarifies that
whenever a given positive random partition $p$ generates the potential for TU
games as the expected accumulated worth of the random partition, then this can
be extended to a potential for TUX games in a unique plausible way. This is
achieved by using the restriction operator $r^{p}$ given in (\ref{eq:r-p}),
which yields the auxiliary games $v_{w}^{r^{p}}$ as defined in
(\ref{eq:w-tilde-r}), thereby yielding the potential $\mathrm{Pot}^{r^{p}}$
according to (\ref{eq:Pot-r-w}).

Moreover, Theorem~\ref{thm:canext} is constructive in the sense that it yields
a tractable formula for the restriction operator $r^{p}$. This has immediate
consequences, such as a simple formula for the induced $r^{p}$-Shapley value,
which by definition is a player's contribution to the $r^{p}$-potential of a
TUX game. To make this insight more precise, we define a new class of
solutions for TUX\ games. Note that the following definition is meaningful for
any random partition.

\begin{definition}
Let $p$ be a random partition. The $p$-Shapley value $\mathrm{Sh}^{p}$ is
given by%
\begin{align}
\mathrm{Sh}_{i}^{p}\left(  w\right)   &  =\sum_{\left(  T,\tau\right)
\in\mathcal{E}\left(  N\setminus\left\{  i\right\}  \right)  }\left(
\vphantom{\sum_{B\in\tau\cup\left\{ \emptyset\right\} }}%
p_{N}\left(  \left\{  T\cup\left\{  i\right\}  \right\}  \cup\tau\right)
w\left(  T\cup\left\{  i\right\}  ,\tau\right)  \right. \nonumber\\
&  \left.  \qquad\qquad\qquad\qquad\qquad\qquad-\frac{t}{n-t}\sum_{B\in
\tau\cup\left\{  \emptyset\right\}  }p_{N}\left(  \left\{  T\right\}  \cup
\tau_{+i\leadsto B}\right)  w\left(  T,\tau_{+i\leadsto B}\right)  \right)
\label{eq:Sh-rp}%
\end{align}
for all $N\subseteq\mathbf{U},~w\in\mathbb{W}\left(  N\right)  ,$ and $i\in
N.$
\end{definition}

It turns out that $r^{p}$-Shapley values have this structure of a $p$-Shapley value.

\begin{corollary}
\label{cor:sh-rp}(i) Let $p$ be a positive random partition (\textbf{POS})
that generates the potential for TU games (\textbf{GEN}), and denote the
corresponding restriction operator according to (\ref{eq:r-p}) by $r^{p}$.
Then, the $r^{p}$-Shapley value $\mathrm{Sh}^{r^{p}}$ as defined by
(\ref{eq:r-Shapleyvalue}) can be computed as in (\ref{eq:Sh-rp}), i.e., we
have $\mathrm{Sh}^{r^{p}}=\mathrm{Sh}^{p}.$

(ii) Moreover, for TU~games, the $r^{p}$-Shapley value $\mathrm{Sh}^{r^{p}}$
coincides with the Shapley value, i.e., $\mathrm{Sh}^{r^{p}}\left(  v\right)
=\mathrm{Sh}\left(  v\right)  $ for all $N\subseteq\mathbf{U},$ $v\in
\mathbb{V}\left(  N\right)  \subseteq\mathbb{W}\left(  N\right)  $.
\end{corollary}

The corollary reconfirms that the $r^{p}$-Shapley value generalizes the
Shapley value for suitable random partitions. Moreover, it provides a formula
that allows us to investigate its properties.

\begin{remark}
\label{rem:POS}Note that \textbf{GEN} essentially applies to fixed $N.$ Since
\textbf{GEN} implies $p_{N}=p_{N}^{\star}$ for $n\leq3$
(Corollary~\ref{cor:consequence-of-gen}), we can do without positivity
(\textbf{POS}) in Theorem~\ref{thm:canext} and Corollary~\ref{cor:sh-rp} for
$\left\vert \mathbf{U}\right\vert \leq3.$ For $\left\vert \mathbf{U}%
\right\vert =4$, the restriction operators in (\ref{eq:r-p}) are well-defined
and the proofs of Theorem~\ref{thm:canext} and Corollary~\ref{cor:sh-rp} run
through smoothly without positivity (\textbf{POS}).
\end{remark}

\subsection{The only $r^{p}$-Shapley value satisfying the null player player
property is the MPW\ solution\label{sec:MPW}}

So far we have motivated the $r^{p}$-Shapley value, which emerges if we want
to generalize the potential from TU games to TUX games in a way that preserves
its interpretation of being an expected accumulated worth of a random
partition. We further identified the specific structure of such $r^{p}%
$-Shapley values.

From (\ref{eq:Sh-rp}), it becomes apparent that the $r^{p}$-Shapley values are
additive in the game. Moreover, every $r^{p}$-Shapley value inherits
efficiency from the Shapley value by definitions~(\ref{eq:r-Shapleyvalue})
and~(\ref{eq:w-tilde-r}). However, it is less clear whether the $r^{p}%
$-Shapley values satisfy standard monotonicity properties or assign zero
payoff to null players. Indeed, the latter is at the heart of our second main
result. We prepare it by investigating the obvious question of what we obtain
if we apply Theorem~\ref{thm:canext} to the Ewens distribution $p^{\star}$
given in (\ref{eq:p-star}).

\begin{lemma}
\label{lem:rpstar}For $p=p^{\star}$, the restriction operator $r^{\star
}:=r^{p^{\star}}$ is given by%
\begin{equation}
w_{-i}^{r^{\star}}\left(  S,\pi\right)  =\frac{1}{n-s}w\left(  S,\pi
_{+i\leadsto\emptyset}\right)  +\sum_{B\in\pi}\frac{b}{n-s}w\left(
S,\pi_{+i\leadsto B}\right)  \label{eq:r-star}%
\end{equation}
for all $N\subseteq\mathbf{U}$, $w\in\mathbb{W}\left(  N\right)  $, $i\in N,$
and $\left(  S,\pi\right)  \in\mathcal{E}\left(  N\setminus\left\{  i\right\}
\right)  $. Moreover, the $r^{\star}$-Shapley value coincides with the
MPW~solution, $\mathrm{Sh}^{r^{\star}}=\mathrm{MPW}$.
\end{lemma}

Lemma~\ref{lem:rpstar} highlights that $p^{\star}$ yields an interesting
restriction and a very familiar generalization of the Shapley value. Moreover,
the restriction operator lends itself to an intuitive interpretation
reminiscent of the Chinese restaurant process. The worth of an embedded
coalition after player~$i$ has left the game computes as if this player would
take a seat next to any outside player equally likely or stay alone (see the
paragraph prior to (\ref{eq:p-con})).

It is well-known that the MPW solution assigns a zero payoff to null players
\citep[Theorem~2]{MSPCWe2007}%
. Specifically, player $i\in N$ is a\textbf{ null player} in TUX game
$w\in\mathbb{W}\left(  N\right)  $ with player set $N\subseteq\mathbf{U}$ if
player $i$'s position in the game does not influence the worth of any embedded
coalition, i.e., if $w\left(  S\cup\left\{  i\right\}  ,\pi\right)  =w\left(
S,\pi_{+i\leadsto B}\right)  $ for all $\left(  S,\pi\right)  \in
\mathcal{E}\left(  N\setminus\left\{  i\right\}  \right)  $ and $B\in\pi
\cup\left\{  \emptyset\right\}  $.\footnote{%
\citet{DuEhKa2010}%
\ call (our) null players \textquotedblleft dummy players of
type~1\textquotedblright.
\citet{bolger1989}%
\ calls them \textquotedblleft dummies\textquotedblright,
\citet{MSPCWe2007}%
\ \textquotedblleft dummy players\textquotedblright, and
\citet{CliSer2008}%
\ \textquotedblleft null players in the strong sense\textquotedblright.} The
MPW\ solution satisfies the following property.\footnote{%
\citet{MSPCWe2007}%
\ call this property the \textquotedblleft dummy player
axiom\textquotedblright. It can be verified as follows. By (\ref{eq:r-star}),
we have $w\left(  S\cup\left\{  i\right\}  ,\pi\right)  =w_{-i}^{r^{\star}%
}\left(  S,\pi\right)  $ for null player $i$; hence, $w_{-N\setminus\left(
S\cup\left\{  i\right\}  \right)  }^{r^{\star}}\left(  S\cup\left\{
i\right\}  ,\emptyset\right)  =w_{-N\setminus S}^{r^{\star}}\left(
S,\emptyset\right)  $, i.e., player $i$ is a null player in $v_{w}^{r}$
(\ref{eq:w-tilde-r}); applying the Shapley value to $v_{w}^{r^{\star}}$ then
allocates zero to player $i,$ $\mathrm{Sh}_{i}\left(  v_{w}^{r^{\star}%
}\right)  =0$; and by Lemma~\ref{lem:rpstar}, $\mathrm{MPW}_{i}\left(
w\right)  =\mathrm{Sh}_{i}^{r^{\star}}\left(  w\right)  =\mathrm{Sh}%
_{i}\left(  v_{w}^{r^{\star}}\right)  =0$.}

\smallskip

\noindent\textbf{Null player, NP.}$\;$For all $N\subseteq\mathbf{U},$
$w\in\mathbb{W}\left(  N\right)  ,$ and $i\in N$ such that player $i$ is a
null player in $w,$ we have $\varphi_{i}\left(  w\right)  =0.$

\smallskip

Notice that the null player property imposes the standard null player property
on TU games $\mathbb{V}\subseteq\mathbb{W}$.\footnote{A solution $\varphi$
satisfies the null player property for TU\ games if for all $v\in
\mathbb{V}\left(  N\right)  \subseteq\mathbb{W}\left(  N\right)  ,$
$N\in\mathcal{N}$, $i\in N$ such that $v\left(  S\cup\left\{  i\right\}
\right)  =v\left(  S\right)  $ for all $S\subseteq N\setminus\left\{
i\right\}  $, we have $\varphi_{i}\left(  v\right)  =0$.} Indeed, since every
$r^{p}$-Shapley value coincides with the Shapley value on TU\ games, every
$r^{p}$-Shapley value also satisfies the null player property for TU~games.
However, the next proposition clarifies that only one $r^{p}$-Shapley value
satisfies the null player property also in the presence of externalities: the
MPW solution. In fact, we show this claim for the larger class of $p$-Shapley
values (with arbitrary random partitions).

\begin{proposition}
\label{thm:rp+null=MPW}Let $p$ be a random partition and denote the
corresponding $p$-Shapley value defined in (\ref{eq:Sh-rp}) by $\mathrm{Sh}%
^{p}$. The $p$-Shapley value $\mathrm{Sh}^{p}$ satisfies the null player
property (\textbf{NP}) if and only if $p=p^{\star}$, i.e., if and only if
$\mathrm{Sh}^{p}=\mathrm{MPW}$.
\end{proposition}

%

\citet{CliSer2008}
put forth another important property for solutions for TUX games.
Specifically, they argue that a player's payoff shall weakly increase with
this player's marginal contributions.

\smallskip

\noindent\textbf{Monotonicity, M.}$\;$For all $N\subseteq\mathbf{U},$
$w,z\in\mathbb{W}\left(  N\right)  $ and $i\in N,$ such that%
\[
w\left(  S\cup\left\{  i\right\}  ,\pi\right)  -w\left(  S,\pi_{+i\leadsto
B}\right)  \geq z\left(  S\cup\left\{  i\right\}  ,\pi\right)  -z\left(
S,\pi_{+i\leadsto B}\right)
\]
for all $\left(  S,\pi\right)  \in\mathcal{E}\left(  N\setminus\left\{
i\right\}  \right)  $ and $B\in\pi\cup\left\{  \emptyset\right\}  ,$ we have
$\varphi_{i}\left(  w\right)  \geq\varphi_{i}\left(  z\right)  .$

\smallskip

It is well-known that the MPW solution satisfies monotonicity
\citep[p.~346]{MSPCWe2007}%
. Moreover, every $p$-Shapley value gives a zero payoff to every player in the
null game, $\mathrm{Sh}_{i}^{p}\left(  \mathbf{0}^{N}\right)  =0$ for all
$i\in N,$ $N\subseteq\mathbf{U}$. Thus, a $p$-Shapley value satisfies the null
player property whenever it is monotonic, because a null player in a TUX\ game
has the same marginal contributions as in the null game. We can therefore
replace the null player property by monotonicity in
Proposition~\ref{thm:rp+null=MPW}.

\begin{corollary}
\label{cor:rp+mo=MPW}Let $p$ be a random partition and let $\mathrm{Sh}^{p}$
denote the corresponding $p$-Shapley value as defined in (\ref{eq:Sh-rp}). The
$p$-Shapley value $\mathrm{Sh}^{p}$ satisfies monotonicity (\textbf{M}) if and
only if $p=p^{\star}$, i.e., if and only if $\mathrm{Sh}^{p}=\mathrm{MPW}$.
\end{corollary}

Finally, we can combine the insights we gathered so far. The \emph{MPW
solution\ emerges as the unique solution }$\varphi$\emph{ for TUX games if} we
demand the following:

\begin{enumerate}
\item the TUX solution $\varphi$ is an $r$-Shapley value for \emph{some}%
\textit{\emph{\/}} path independent (\textbf{PI}) restriction operator $r$
that preserves null games (\textbf{PNG});

\item its corresponding $r$-potential coincides with the expected accumulated
worth for \emph{some}\textit{\emph{\/}} random partition $p$ that generates
the potential for TU games (\textbf{GEN}); and

\item the TUX solution $\varphi$ satisfies the null player property
(\textbf{NP}) or monotonicity (\textbf{M}).
\end{enumerate}

This statement, in particular the fact that we can omit the positivity
condition (\textbf{POS}) on the random partition $p$, is clarified in our
final theorem.

\begin{theorem}
\label{thm:MPW}The random partition $p^{\star}$ is the unique random partition
$p$ with the following properties:

(i) The random partition $p$ generates the potential for TU games
(\textbf{GEN}).

(ii) There exists a path-independent (\textbf{PI}) restriction operator $r$
that preserves null games (\textbf{PNG}), that satisfies $\mathrm{Pot}%
^{r}=\mathrm{E}_{p}$, i.e., its induced $r$-potential defined in
(\ref{eq:Pot-r-w}) coincides with the expected accumulated worth of the random
partition $p$ given in (\ref{eq:def-Ep}), and such that the corresponding
$r$-Shapley value $\mathrm{Sh}^{r}$ as defined in (\ref{eq:r-Shapleyvalue})
satisfies the [null player property (\textbf{NP}) or monotonicity (\textbf{M})].

Moreover, the restriction operator $r^{\ast}$ given in (\ref{eq:r-star}) is
the unique such restriction operator, which induces the MPW solution,
$\mathrm{Sh}^{r^{\star}}=\mathrm{MPW}.$
\end{theorem}

We conclude this subsection with a discussion of the role of positivity
(\textbf{POS}) in our paper. In Remark~\ref{rem:POS}, we already have provided
an argument why we do not need positivity in Theorem~\ref{thm:canext} and
Corollary~\ref{cor:sh-rp} for $\left\vert \mathbf{U}\right\vert \leq4$. It may
seem that positivity is \textquotedblleft only\textquotedblright\ needed to
make sure that the restriction operators in (\ref{eq:r-p}) are
\emph{well-defined}. Actually, positivity is also used to ensure the
\emph{existence}\textit{\emph{\/}} of the restriction operators as in
Theorem~\ref{thm:canext}. Indeed, it remains an open question whether such
restriction operators actually do exist for random partitions that fail
positivity for $\left\vert \mathbf{U}\right\vert >4$.

This might raise the question of why positivity is not required in
Theorem~\ref{thm:MPW}. This can best been seen from a sketch of its proof.
Property \textbf{GEN} (generation of the potential for TU games) implies
$p_{N}=p_{N}^{\star}\ $for $n\leq3~$(Corollary~\ref{cor:consequence-of-gen}).
Now, we proceed by induction on $n.$ Assume $p_{N}=p_{N}^{\star}$ for $n\leq
a$ and consider $N$ with $n=a+1.$ (i) Since the random partition $p^{\star}$
satisfies positivity (\textbf{POS}), the restriction operator $r^{p}$ is
well-defined for $N.$ (ii) Arguments as in its proof show that $r^{p}$ is the
unique restriction operator as in Theorem~\ref{thm:canext} for $N.$ (iii)
Arguments as in the proof of Corollary~\ref{cor:sh-rp} show that the
restriction operator $r^{p}$ induces the $p$-Shapley value for $N.$ (iv)
Arguments as in the proof of Theorem~\ref{thm:rp+null=MPW} or
Corollary~\ref{cor:rp+mo=MPW}, respectively, show that the null player
property (\textbf{NP}) or monotonicity (\textbf{M}) imply $p_{N}=p_{N}^{\star
}.$ This already implies $p_{N}=p_{N}^{\star}$ for all $N\subseteq\mathbf{U}.$
The rest of the claim now follows from Theorem~\ref{thm:Pot-p-star},
Theorem~\ref{thm:canext}, Corollary~\ref{cor:sh-rp}, and
Lemma~\ref{lem:rpstar}.

\section{Discussion}

In the previous section, we have deduced the MPW\ solution with some
regularity conditions as the unique TUX solution that satisfies the null
player property, and that admits a potential for TUX games that can be
computed as the expected accumulated worth of a random partition. In this
section, we connect our approach with the conditional independence property of
random partitions
\citep{kingman1978}%
, which is used by
\citet{SkiMic2020}
to characterize the MPW solution. Moreover, we relate our findings to other
solution concepts mentioned in the literature. Finally, we discuss\ the
preservation of null players property for restriction operators, which is
emphasized by
\citep{DuEhKa2010}%
\ but violated by the restriction operator of the MPW solution.

\subsection{Potential generation and conditional independence\label{sec:CI}}

We relate the condition on a random partition $p$ that generates the potential
for TU\ games (\textbf{GEN}) to another prominent condition on random
partitions, which is known as conditional independence.

\smallskip

\noindent\textbf{Conditional independence, CI}
\citep{kingman1978}%
\textbf{.}$\;$For all $N\subseteq\mathbf{U},~\pi\in\Pi\left(  N\right)  ,$ and
$B\in\pi$, we have%
\begin{equation}
p_{N}\left(  \pi\right)  =p_{N\setminus B}\left(  \pi\setminus\left\{
B\right\}  \right)  \sum_{\tau\in\Pi\left(  N\right)  :B\in\tau}p_{N}\left(
\tau\right)  \text{.} \label{eq:CI}%
\end{equation}

With some mild additional assumptions, conditional independence is
characteristic of the family of the Ewens distributions. Note that conditional
independence does not imply potential generation, nor the other way
around.\footnote{The Ewens distributions defined in Footnote~\ref{fn:ewens}
with mutation rates $\theta\neq1$ are known to satisfy \textbf{CI}, but
violate \textbf{GEN}, e.g., $p_{N}^{1/2}\left(  \left\{  N\right\}  \right)
\neq1/n$---a contradiction to \ref{cor:consequence-of-gen}. The example in
(\ref{eq:pot4players}) with $c=1/8$ satisfies \textbf{GEN}, but violates
\textbf{CI}, which can be seen from $p_{4}^{1/8}\left(  \left\{  \left\{
i,j\right\}  ,\left\{  k\right\}  ,\left\{  \ell\right\}  \right\}  \right)
=0\neq p_{2}^{\star}\left(  \left\{  \left\{  k\right\}  ,\left\{
\ell\right\}  \right\}  \right)  \cdot1/12$, a contradiction to (\ref{eq:CI})
for $B=\left\{  i,j\right\}  $.} In this sense, our work complements
\citet{SkiMic2020}%
, who highlight how conditional independence selects the MPW solution among
those solutions for TUX games that can be motivated via a stochastic partition
formation process. However, only one random partition satisfies both conditions.

\begin{proposition}
\label{pro:CI+GEN=/p-star}The random partition $p^{\star}$ given in
(\ref{eq:p-star}) is the unique random partition that generates the potential
(\textbf{GEN}) and satisfies conditional independence (\textbf{CI}).
\end{proposition}

\subsection{Other prominent TUX solutions are no $p$-Shapley
values\label{sec:otherTUXsolutions}}

In Corollary \ref{cor:sh-rp}, we essentially find that whenever a TUX solution
admits a potential that can be computed as the expected accumulated worth of a
positive random partition~$p$, then it belongs to the class of $p$-Shapley
value. In view of the uniqueness result of Theorem \ref{thm:MPW}, it might not
surprise that other prominent TUX solutions do not belong to this class. In
particular, the TUX solutions suggested by
\citet{myerson1977pffg}%
,
\citet{PhaNor2007}%
,\ and
\citet[Definition~4]{mcquillin2009}%
\ coincide with the MPW solution for $n\leq2$ but are not $p$-Shapley values
for $n>2.$ The TUX solutions suggested by
\citet{bolger1989}%
\ and
\citet{AlArRu2005}%
\ coincide with the MPW solution for $n\leq3$ but are not $p$-Shapley values
for $n>3.$ The class of solutions identified by
\citet[Theorem~2]{DuEhKa2010}
does not contain the MPW solution, but each of its members satisfies the null
player property. Hence, it does not intersect with the $p$-Shapley
values.\footnote{These relationships can be obtained by the straightforward
but tedious calculation of the payoffs for Dirac games. Details can obtained
from the authors upon request.}

\subsection{Null players after removing another player\label{sec:NP}}

Although the MPW\ solution is obtained as an intuitive generalization of the
Shapley value that can be interpreted as a player's contribution to the
expected accumulated worth of a random partition in a game, its corresponding
restriction operator, i.e., its notion of how to obtain subgames, exhibits a
surprising property. Suppose player~$j$ is a null player in a game and some
other player~$i$ is removed from the game. It may happen that player~$j$ is no
longer a null player in the subgame. Note, however, that player~$j$'s payoff
according to the MPW solution remains zero. To see this more lucidly, consider
the following example.

\begin{example}
There are four players, $N=\left\{  1,2,3,4\right\}  .$ The partition function
$w$ is given by%
\begin{equation}
w\left(  S,\pi\right)  =\left\{
\begin{array}
[c]{ll}%
1, & \text{if }2\in S\text{ and }3\in S,\\
1, & \text{if }2\in S,\text{ }3\notin S,\text{ and }4\notin\pi\left(
3\right)  ,\\
1, & \text{if }3\in S,\text{ }2\notin S,\text{ and }4\notin\pi\left(
2\right)  ,\\
0, & \text{otherwise.}%
\end{array}
\right.  \qquad\text{for all }\left(  S,\pi\right)  \in\mathcal{E}\left(
N\right)  . \label{eq:exa-w}%
\end{equation}
In words, the position of player~$1$ is irrelevant. Players~$2$ and~$3$ are
the productive players. If both of them are contained in $S,$ then a worth of
$1$ is created. Yet, if only one of them is contained in $S,$ then the worth
created depends on the position of player $4.$ If player $4$ is in the same
block as the productive player not in $S,$ then the worth created drops to
$0.$
\end{example}

In this game, player~$1$ is a null player, whereas player~$4$'s impact
originates from external effects only. Assume that player~$4$ leaves the game.
In the resulting subgame, the worth $w_{-4}^{r}\left(  \left\{  1,3\right\}
,\left\{  \left\{  2\right\}  \right\}  \right)  $ is determined by
\[
w\left(  \left\{  1,3\right\}  ,\left\{  \left\{  2\right\}  ,\left\{
4\right\}  \right\}  \right)  =1\quad\text{and\quad}w\left(  \left\{
1,3\right\}  ,\left\{  \left\{  2,4\right\}  \right\}  \right)  =0,
\]
and the worth $w_{-4}^{r}\left(  \left\{  3\right\}  ,\left\{  \left\{
1,2\right\}  \right\}  \right)  $ is determined by%
\[
w\left(  \left\{  3\right\}  ,\left\{  \left\{  1,2,4\right\}  \right\}
\right)  =0\quad\text{and\quad}w\left(  \left\{  3\right\}  ,\left\{  \left\{
1,2\right\}  ,\left\{  4\right\}  \right\}  \right)  =1.
\]
Thus, player $1$'s marginal contribution $w_{-4}^{r}\left(  \left\{
1,3\right\}  ,\left\{  \left\{  2\right\}  \right\}  \right)  -w_{-4}%
^{r}\left(  \left\{  3\right\}  ,\left\{  \left\{  1,2\right\}  \right\}
\right)  $ depends on these numbers. A restriction operator needs to aggregate
this information and process the externalities exhibited by player~$4$. The
restriction operator corresponding to the MPW solution $r^{\star}$ preserves
the fact that not only players~$2$ and~$3$ have an influence on creation of
worth in the original game, and make player~$1$ a non-null player in the
subgame, resulting in the marginal contribution%
\[
w_{-4}^{r^{\star}}\left(  \left\{  1,3\right\}  ,\left\{  \left\{  2\right\}
\right\}  \right)  -w_{-4}^{r^{\star}}\left(  \left\{  3\right\}  ,\left\{
\left\{  1,2\right\}  \right\}  \right)  =\frac{1}{6}.
\]
This relates to the idea that player~$1$'s position might have had an
influence on how likely player~$4$ \textquotedblleft merged\textquotedblright%
\ with player~$2$. In contrast,
\citet{DuEhKa2010}
focus their analysis on restriction operators that preserve null players as
such when other players are removed.

Although player~$1$ no longer is a null player in the subgame $w_{-4}%
^{r^{\star}}$, the MPW solution still assigns a zero payoff to this player,
$\mathrm{MPW}_{1}\left(  w_{-4}^{r^{\star}}\right)  =0$. This is no
coincidence. As the next proposition states, null players continue to obtain a
zero payoff according to the MPW solution in subgames obtained from removing
other players.

\begin{proposition}
\label{pro:w-bar-star-strong-null}For all $N\subseteq\mathbf{U}$,
$w\in\mathbb{W}\left(  N\right)  ,$ $j\in N,$ and $T\subseteq N\setminus
\left\{  j\right\}  $ such that $j$ is a null player in $w,$ we have
$\mathrm{MPW}_{j}\left(  w_{-T}^{r^{\star}}\right)  =0.$
\end{proposition}

\noindent This result rests on the fact that the average operator $\bar
{v}^{\star}$ which is definitive of the MPW solution, the restriction operator
$r^{\star}$, and the restriction of TU games \textquotedblleft work
together\textquotedblright\ properly, i.e., they commute.

\begin{remark}
\label{rem:np-independent-of-presnp}The preservation of null players by a
restriction operator seems to be closely related to the null player property
(\textbf{NP}) of the corresponding solution. However, both properties are
independent. On the one hand, the MPW solution satisfies the null player
property. On the other hand, path independent restriction operators that
preserve null players may induce TUX solutions via (\ref{eq:r-Shapleyvalue})
that fail the null player property. Consider for instance the nullifying
restriction operator $r^{z}$ given by $w_{-i}^{r^{z}}=\mathbf{0}%
^{N\setminus\left\{  i\right\}  }$ for all $w\in\mathbb{W}\left(  N\right)  ,$
$N\subseteq\mathbf{U}$, and $i\in N.$ Obviously, this is a path independent
restriction operator that preserves null games. Moreover, one easily checks
that the induced TUX solution is the egalitarian TUX solution, $\mathrm{Eg},$
given by $\mathrm{Eg}_{i}\left(  w\right)  =w\left(  N,\emptyset\right)  /n$
for all $w\in\mathbb{W}\left(  N\right)  ,$ $N\subseteq\mathbf{U}$ and $i\in
N,$ which fails the null player property for $n>1$.
\end{remark}

\section{Conclusion}

In this paper, we extend the stream of research on coalitional games with
externalities (TUX games) and varying player sets. We start with the
observation that the Shapley value is the contribution to the potential of a
game, which can be computed as the expected accumulated worth of a random
partition of the player set. In our quest to generalize this idea to TUX
games, we need to address the challenge that there is no obvious way for
creating subgames in the presence of externalities. To this end, we use the
concept of restriction operators introduced by
\citet{DuEhKa2010}%
, which capture a plethora of possibilities to obtain subgames. Each path
independent restriction operator induces a potential for TUX games. Our first
main result describes those restriction operators for which the corresponding
potential can be computed as an expected accumulated worth of some random
partition and which generalize the potential for TU\ games. Each such
restriction operator then induces a generalization of the Shapley value for
TUX games. In particular, each such solution for TUX games gives a payoff zero
to null players in the absence of externalities. However, only the MPW
solution
\citep{MSPCWe2007}
maintains this property also in the presence of externalities, which is our
second main result. We further find that the notion of subgames that
corresponds to the MPW solution relates to the \textquotedblleft Chinese
restaurant process\textquotedblright.

The research on varying player sets not only complements the research on TUX
games that operates on a fixed player set
\citep
{myerson1977pffg,CliSer2008,GraFun2012,sanchezperez2017,SkMiWo2018,YaSuHoXu2019,SkiMic2020}%
, but has multiple motivations as it also paves the way for further research.
For instance, in the context of TU games,
\citet{myerson1980}
uses the removal of a player to capture a notion of fairness: player$~i$ shall
benefit from player$~j$'s participation in the games as much as the other way
around. Within the realm of TU games, this is as characteristic of the Shapley
value as the admittance of a potential function (see
\cite{CasHue-deva}
for a survey of equivalent properties and
\citet{KamKon2012}
for variants that include aspects of solidarity). Analogous fairness
properties can be studied for TUX games based on the intuitive restriction
operator that we introduce.

Another stream of research involves the study of nonlinear generalizations of
the Shapley value to TUX games. Whereas
\citet{young1985}%
\ demonstrates that monotonicity together with efficiency and symmetry are
characteristic of the Shapley value,
\citet{CliSer2008}%
\ emphasize that imposing these axioms in the presence of externalities does
not achieve uniqueness, but even allows for nonlinear solutions.
\citet{DuEhKa2010}%
\ demonstrate that nonlinear solutions are compatible with restriction
operators and their potential approach. This invokes the question of how our
formula of the potential as an expected value of a random partition needs to
be adjusted to allow for such solutions as well.

Moreover, varying player sets are required for the Nash program, which aims at
connecting cooperative and non-cooperative game theory. Most implementations
of the Shapley value make direct or implicit use of reduced games as for
example
\citet{KriSer1995}%
,
\citet{HarMas1996}%
, {%
\citet{gul1989}%
}, {%
\citet{StoZwi1996}%
}, {%
\citet{PerWet2001}%
}, {%
\citet{McQSug2016}%
}, {%
\citet{BruGauMen2018}%
}. The present paper links the restriction operator that is characteristic of
the MPW~solution to the uniform Chinese restaurant process, indicating a
building block for possible implementations of the MPW\ solution.

\section{Acknowledgements}

We would like to thank the associate editor and three referees for pushing us
to substantially improve our paper. Andr\'{e} Casajus: Funded by the Deutsche
Forschungsgemeinschaft (DFG, German Research Foundation) -- 522837108.
Yukihiko Funaki: This work was supported by the JSPS Core-to-Core Program
(JPJSCCA20200001), and JSPS KAKENHI Grant Numbers JP22H00829 and JP18KK0046.

\appendix

\section{Additional notation\label{apx:addnotation}}

We denote the block of $\pi\in\Pi\left(  N\right)  $ that contains player
$i\in N$ by $\pi\left(  i\right)  $. The atomistic partition is denoted by
$\left[  N\right]  \in\Pi\left(  N\right)  $ and given by $\left[  N\right]
:=\left\{  \left\{  i\right\}  \mid i\in N\right\}  .$ For $\pi\in\Pi\left(
N\right)  ,$ $N\subseteq\mathbf{U},$ the elimination of the players in
$T\subseteq N$ from $\pi$ gives $\pi_{-T}\in\Pi\left(  N\setminus T\right)  ,$%
\[
\pi_{-T}:=\left\{  B\setminus T\mid B\in\pi\text{ and }B\setminus
T\neq\emptyset\right\}  .
\]
Instead of $\pi_{-\left\{  i\right\}  }$, we write $\pi_{-i}.$

For $\alpha\in\mathbb{R}$ and $w,w^{\prime}\in\mathbb{W}\left(  N\right)  ,$
$N\subseteq\mathbf{U,}$ the games $\alpha w\in\mathbb{W}\left(  N\right)  $
and $w+w^{\prime}\in\mathbb{W}\left(  N\right)  $ are given by $\left(  \alpha
w\right)  \left(  S,\pi\right)  =\alpha\left(  w\left(  S,\pi\right)  \right)
$ and $\left(  w+w^{\prime}\right)  \left(  S,\pi\right)  =w\left(
S,\pi\right)  +w^{\prime}\left(  S,\pi\right)  $ for all $\left(
S,\pi\right)  \in\mathcal{E}\left(  N\right)  .$ For $N\subseteq\mathbf{U}$
and $\left(  T,\tau\right)  \in\mathcal{E}\left(  N\right)  ,$ $T\neq
\emptyset,$ the \textbf{Dirac (TUX)~game} $\delta_{T,\tau}\in\mathbb{W}\left(
N\right)  $ is given by%
\begin{equation}
\delta_{T,\tau}\left(  S,\pi\right)  :=\left\{
\begin{array}
[c]{ll}%
1, & \text{if }\left(  S,\pi\right)  =\left(  T,\tau\right)  ,\\
0, & \text{else}%
\end{array}
\right.  \qquad\text{for all }\left(  S,\pi\right)  \in\mathcal{E}\left(
N\right)  . \label{eq:dirac}%
\end{equation}
Every TUX game $w\in\mathbb{W}\left(  N\right)  $, $N\subseteq\mathbf{U}$ has
a unique linear representation in terms of Dirac games,%
\begin{equation}
w=\sum_{\left(  T,\tau\right)  \in\mathcal{E}\left(  N\right)  :T\neq
\emptyset}w\left(  T,\tau\right)  \delta_{T,\tau}. \label{eq:dirac-rep}%
\end{equation}
That is, the set $\left\{  \delta_{T,\tau}\mid\left(  T,\tau\right)
\in\mathcal{E}\left(  N\right)  :T\neq\emptyset\right\}  $ is a basis of the
vector space $\mathbb{W}\left(  N\right)  $.

For $N\subseteq\mathbf{U}$ and $T\subseteq N,$ $T\neq\emptyset,$ the
\textbf{Dirac TU~game} $\delta_{T}^{N}\in\mathbb{V}\left(  N\right)  $ is
given by $\delta_{T}^{N}\left(  S\right)  =1$ for $T=S$ and $\delta_{T}%
^{N}\left(  S\right)  =0$ otherwise (we carry the superscript indicating the
player set for such TU games). For $N\subseteq\mathbf{U}$ and $T\subseteq N,$
$T\neq\emptyset,$ the \textbf{unanimity TU~game} $u_{T}^{N}\in\mathbb{V}%
\left(  N\right)  $ is given by $u_{T}^{N}\left(  S\right)  =1$ for
$T\subseteq S$ and $u_{T}^{N}\left(  S\right)  =0$ otherwise.

\section{Proofs}

We now present the proofs of our results.

\subsection{Proof of Proposition~\ref{pro:p=/p-star}}

Both, the mappings $\mathrm{Pot}$ and $\mathrm{E}_{p}$ are linear. Hence, it
suffices to show the claim for Dirac TU games.

(i) $\Rightarrow$ (ii) From (\ref{eq:Pot2}), we obtain the potential for the
Dirac TU games, $\delta_{T}^{N}$, $T\subseteq N,$ $T\neq\emptyset
,\ N\subseteq\mathbf{U}$,
\begin{equation}
\mathrm{Pot}\left(  \delta_{T}^{N}\right)  =\dfrac{\left(  n-t\right)
!\left(  t-1\right)  !}{n!}. \label{eq:pot-TU-dirac}%
\end{equation}
By \textbf{GEN}, we further have $\mathrm{Pot}\left(  \delta_{T}^{N}\right)
=\mathrm{E}_{p}\left(  \delta_{T}^{N}\right)  $. From the definition of
$\mathrm{E}_{p}$, we get%
\[
\mathrm{E}_{p}\left(  \delta_{T}^{N}\right)  =\sum_{\pi\in\Pi\left(  N\right)
:\pi\ni T}p_{N}\left(  \pi\right)  .
\]
This confirms the claim.

(ii) $\Rightarrow$ (iii) The RHS of (\ref{eq:p-reduction}) can be written as
$\frac{n}{n-s}\sum_{\pi\in\Pi\left(  N\setminus S\right)  }p_{N}\left(
\left\{  S\right\}  \cup\pi\right)  $. The claim now follows from (ii).

(iii) $\Rightarrow$ (i) We proceed by induction on $n$.

\emph{Induction basis:}\textit{\emph{\/ }}For $n=1$, $\sum_{\pi\in\Pi\left(
N\right)  }p_{N}\left(  \pi\right)  =1$ implies the claim.

\emph{Induction hypothesis (IH):}\textit{\emph{\/ }}For all $N\subseteq
\mathbf{U}$, $v\in\mathbb{V}\left(  N\right)  $ such that $n<m$, we have
$\mathrm{E}_{p}\left(  v\right)  =\mathrm{Pot}\left(  v\right)  $.

\emph{Induction step:}\textit{\emph{\/ }}Now let $N\subseteq\mathbf{U}$ be
such that $n=m$. For $T\subsetneq N$, $i\in N\setminus T$ we have
\begin{align*}
\mathrm{E}_{p}\left(  \delta_{T}^{N}\right)   &  \overset
{\text{(\ref{eq:def-Ep})}}{=}\sum_{\pi\in\Pi\left(  N\setminus T\right)
}p_{N}\left(  \left\{  T\right\}  \cup\pi\right) \\
&  \overset{\text{(iii)}}{=}\frac{n-t}{n}\cdot\sum_{\tau\in\Pi\left(  \left(
N\setminus\left\{  i\right\}  \right)  \setminus T\right)  }p_{N\setminus
\left\{  i\right\}  }\left(  \left\{  T\right\}  \cup\tau\right) \\
&  \overset{\text{(\ref{eq:def-Ep})}}{=}\frac{n-t}{n}\cdot\mathrm{E}%
_{p}\left(  \delta_{T}^{N\setminus\left\{  i\right\}  }\right) \\
&  \overset{\text{\emph{IH}}}{=}\frac{n-t}{n}\cdot\mathrm{Pot}\left(
\delta_{T}^{N\setminus\left\{  i\right\}  }\right) \\
&  \overset{\text{(\ref{eq:pot-TU-dirac})}}{=}\frac{n-t}{n}\cdot\frac{\left(
n-t-1\right)  !\left(  t-1\right)  !}{\left(  n-1\right)  !}\\
&  \overset{\text{(\ref{eq:pot-TU-dirac})}}{=}\mathrm{Pot}\left(  \delta
_{T}^{N}\right)  ,
\end{align*}
where we use the definition (\ref{eq:def-Ep}) of $\mathrm{E}_{p}$ in the first
and third equation, condition (iii) in the second equation, the induction
hypothesis in the fourth equation, and the formula for the potential of
TU\ Dirac games (\ref{eq:pot-TU-dirac}) in the last equation.

Next we deal with $\delta_{T}^{N}$. To this end, fix some player $i\in N$.
Summing up $\sum_{T\subsetneq N:i\in T}\mathrm{Pot}\left(  \delta_{T}%
^{N}\right)  =\sum_{\pi\in\Pi\left(  N\setminus T\right)  }p_{N}\left(
\left\{  T\right\}  \cup\pi\right)  $ for all $T\subsetneq N$ such that $i\in
T$ gives%
\[
\sum_{T\subsetneq N:i\in T}\mathrm{Pot}\left(  \delta_{T}^{N}\right)
=\sum_{T\subsetneq N:i\in T}\sum_{\tau\in\Pi\left(  N\setminus T\right)
}p_{N}\left(  \left\{  T\right\}  \cup\tau\right)  =\sum_{\pi\in\Pi\left(
N\right)  }p_{N}\left(  \pi\right)  -p_{N}\left(  \left\{  N\right\}  \right)
.
\]
Note that $\sum_{T\subsetneq N:i\in T}\mathrm{Pot}\left(  \delta_{T}%
^{N}\right)  =\mathrm{Pot}\left(  u_{\left\{  i\right\}  }^{N}-\delta_{N}%
^{N}\right)  =1-\frac{1}{n}$. On the other hand, we have $\sum_{\pi\in
\Pi\left(  N\right)  }p_{N}\left(  \pi\right)  -p_{N}\left(  \left\{
N\right\}  \right)  =1-p_{N}\left(  \left\{  N\right\}  \right)  $. Hence,
$p_{N}\left(  \left\{  N\right\}  \right)  =\frac{1}{n}$, and therefore
$\mathrm{E}_{p}\left(  \delta_{N}^{N}\right)  =p_{N}\left(  \left\{
N\right\}  \right)  =\mathrm{Pot}\left(  \delta_{N}^{N}\right)  $.

Thus, we have $\mathrm{E}_{p}\left(  \delta_{T}^{N}\right)  =\mathrm{Pot}%
\left(  \delta_{T}^{N}\right)  $ for all $T\subseteq N$. Note that the last
equation is linear in the game $v=\sum_{T\neq\emptyset}v\left(  T\right)
\delta_{T}^{N}$, so that
\[
\mathrm{E}_{p}\left(  v\right)  =\sum_{T\subseteq N:T\neq\emptyset}v\left(
T\right)  \mathrm{E}_{p}\left(  \delta_{T}^{N}\right)  =\sum_{T\subseteq
N:T\neq\emptyset}v\left(  T\right)  \mathrm{Pot}\left(  \delta_{T}^{N}\right)
=\mathrm{Pot}\left(  v\right)
\]
for all $v\in\mathbb{V}\left(  N\right)  $.

\subsection{Proof of Corollary~\ref{cor:consequence-of-gen}}

$\left(  i\right)  $ This immediately follows from setting $\pi=\left\{
N\right\}  $ and $\pi=\left\{  N\setminus\left\{  i\right\}  ,\left\{
i\right\}  \right\}  $ in (\ref{eq:prop-p-gives-Pot-for-Dirac}), respectively.

$\left(  ii\right)  $ This immediately follows from $\left(  i\right)  $ and
$p_{N}$ being a probability distribution on $\Pi\left(  N\right)  .$

\subsection{Generalization of the example with four players in
(\ref{eq:pot4players}) to larger\ games}

There exist random partitions that satisfy \textbf{GEN} but deviate from
$p^{\star}$ for arbitrary $n\geq4$. Let $\mathbf{U}_{>3}:=\left\{
4,5,6,\dots\right\}  $ and $\varepsilon:\mathbf{U}_{>3}\rightarrow\mathbb{R},$
$k\mapsto\varepsilon_{k}$ be such that $\varepsilon_{k}\in\left[  -\frac
{1}{k!},\frac{\binom{k}{2}}{2\cdot k!}\right]  .$ Consider the random
partitions $p^{\varepsilon}$ given by%
\begin{equation}
p_{N}^{\varepsilon}\left(  \pi\right)  =\left\{
\begin{array}
[c]{ll}%
p_{N}^{\star}\left(  \pi\right)  +\frac{2}{\binom{n-2}{2}\binom{n}{2}%
}\varepsilon_{n}, & n>3,~\pi=\left\{  \left\{  i,j\right\}  ,\left\{
k,\ell\right\}  \right\}  \cup\left[  N\setminus\left\{  i,j,k,\ell\right\}
\right]  ,\\
p_{N}^{\star}\left(  \pi\right)  -\frac{2}{\binom{n}{2}}\varepsilon_{n}, &
n>3,~\pi=\left\{  \left\{  i,j\right\}  \right\}  \cup\left[  N\setminus
\left\{  i,j\right\}  \right]  ,\\
p_{N}^{\star}\left(  \pi\right)  +\varepsilon_{n}, & n>3,~\pi=\left[
N\right]  ,\\
p_{N}^{\star}\left(  \pi\right)  , & \text{otherwise}%
\end{array}
\right.  \label{eq:p-epsilon}%
\end{equation}
for all $N\subseteq\mathbf{U},$ pairwise different $i,j,k,\ell\in N,$ and
$\pi\in\Pi\left(  N\right)  $. Straightforward but tedious calculations show
that these random partitions satisfy \textbf{GEN}. Details can obtained from
the authors upon request.

\subsection{Proof of Theorem~\ref{thm:canext}}

Let $p$ be a positive random partition (\textbf{POS}) that satisfies
\textbf{GEN}.

\emph{Existence:}\textit{\emph{\/}} The restriction operator $r^{p}$ given by
(\ref{eq:r-p}) obviously satisfies \textbf{PNG}. We next investigate the
$r^{p}$-restriction of TUX Dirac games. For all $N\subseteq\mathbf{U}$, $i\in
N,$ $\left(  T,\tau\right)  \in\mathcal{E}\left(  N\right)  ,$ $T\neq
\emptyset$, $\alpha\in\mathbb{R},$ and $\left(  S,\pi\right)  \in
\mathcal{E}\left(  N\setminus\left\{  i\right\}  \right)  ,$ we obtain%
\begin{align*}
\left(  \alpha\delta_{T,\tau}\right)  _{-i}^{r}\left(  S,\pi\right)   &
\overset{\text{(\ref{eq:r-p})}}{=}\frac{n}{n-s}\sum_{B\in\pi\cup\left\{
\emptyset\right\}  }\frac{p_{N}\left(  \left\{  S\right\}  \cup\pi_{+i\leadsto
B}\right)  }{p_{N\setminus\left\{  i\right\}  }\left(  \left\{  S\right\}
\cup\pi\right)  }\alpha\delta_{T,\tau}\left(  S,\pi_{+i\leadsto B}\right) \\
&  \overset{\text{(\ref{eq:dirac})}}{=}\sum_{B\in\pi\cup\left\{
\emptyset\right\}  }\left\{
\begin{array}
[c]{ll}%
\alpha\frac{n}{n-s}\frac{p_{N}\left(  \left\{  S\right\}  \cup\pi_{+i\leadsto
B}\right)  }{p_{N\setminus\left\{  i\right\}  }\left(  \left\{  S\right\}
\cup\pi\right)  }, & \text{if }\left(  S,\pi_{+i\leadsto B}\right)  =\left(
T,\tau\right)  ,\\
0, & \text{if }\left(  S,\pi_{+i\leadsto B}\right)  \neq\left(  T,\tau\right)
\end{array}
\right. \\
&  =\left\{
\begin{array}
[c]{ll}%
\alpha\frac{n}{n-t}\frac{p_{N}\left(  \left\{  T\right\}  \cup\tau\right)
}{p_{N\setminus\left\{  i\right\}  }\left(  \left\{  T\right\}  \cup\tau
_{-i}\right)  }, & \text{if }i\in N\setminus T\text{ and }\left(
S,\pi\right)  =\left(  T,\tau_{-i}\right)  ,\\
0, & \text{otherwise,}%
\end{array}
\right.
\end{align*}
where we use the definition of the restriction operator $r^{p}$ (\ref{eq:r-p})
in the first equation and the definition of TUX Dirac games (\ref{eq:dirac})
in the second equation. Therefore, we have%
\begin{equation}
\left(  \alpha\delta_{T,\tau}\right)  _{-i}^{r^{p}}\overset
{\text{(\ref{eq:dirac})}}{=}\left\{
\begin{array}
[c]{ll}%
\alpha\frac{n}{n-t}\frac{p_{N}\left(  \left\{  T\right\}  \cup\tau\right)
}{p_{N\setminus\left\{  i\right\}  }\left(  \left\{  T\right\}  \cup\tau
_{-i}\right)  }\delta_{T,\tau_{-i}}, & \text{if }i\in N\setminus T,\\
\mathbf{0}^{N\setminus\left\{  i\right\}  }, & \text{if }i\in T\text{.}%
\end{array}
\right.  \label{eq:dirac-r-p}%
\end{equation}

Next, we show that $r^{p}$ is path independent (\textbf{PI}). By the linearity
of $r^{p}$ and since the Dirac games form a basis of the $\mathbb{W}\left(
N\right)  $ (see (\ref{eq:dirac-rep})), it suffices to show path independence
for the Dirac games $\delta_{T,\tau}$. We obtain
\begin{align*}
(\left(  \alpha\delta_{T,\tau}\right)  _{-i}^{r^{p}})_{-j}^{r^{p}}  &
\overset{\text{(\ref{eq:dirac-r-p})}}{=}\left\{
\begin{array}
[c]{ll}%
\alpha\frac{n}{n-t}\frac{n-1}{n-t-1}\frac{p_{N}\left(  \left\{  T\right\}
\cup\tau\right)  }{p_{N\setminus\left\{  i.j\right\}  }\left(  T\cup
\tau_{-\left\{  i,j\right\}  }\right)  }\delta_{T,\tau_{-\left\{  i,j\right\}
}}, & \text{if }i,j\in N\setminus T,\\
\mathbf{0}^{N\setminus\left\{  i,j\right\}  }, & \text{if }i\in T\text{ or
}j\in T
\end{array}
\right. \\
&  \overset{\text{(\ref{eq:dirac-r-p})}}{=}\left(  \left(  \delta_{T,\tau
}\right)  _{-j}^{r^{p}}\right)  _{-i}^{r^{p}}%
\end{align*}
for all $N\subseteq\mathbf{U}$, $\left(  T,\tau\right)  \in\mathcal{E}\left(
N\right)  ,$ and $i,j\in N,$ $i\neq j,$ which establishes \textbf{PI}. We can
therefore remove multiple players in a well-defined manner. In particular, we
have%
\begin{equation}
\left(  \delta_{T,\tau}\right)  _{-S}^{r^{p}}\overset
{\text{(\ref{eq:dirac-r-p})}}{=}\prod_{k=0}^{s-1}\frac{n-k}{n-t-k}\frac
{p_{N}\left(  \left\{  T\right\}  \cup\tau\right)  }{p_{N\setminus S}\left(
\left\{  T\right\}  \cup\tau_{-S}\right)  }\delta_{T,\tau_{-S}}
\label{eq:dirac-r-p-S}%
\end{equation}
for all $N\in\mathcal{N}$, $\left(  T,\tau\right)  \in\mathcal{E}\left(
N\right)  ,$ and $S\in N\setminus T.$

Finally, we show $\mathrm{Pot}^{r^{p}}=\mathrm{E}_{p}.$ By \textbf{PI} and
Theorem$~$(\ref{thm:exist-PF-pot}) due to
\citet{DuEhKa2010}%
, there exists a unique $r^{p}$-potential $\mathrm{Pot}^{r^{p}}.$ Since
$r^{p}$ is linear in the game and by definition of $v_{w}^{r}$
(\ref{eq:w-tilde-r}) as well as the definition of the $r$-potential
(\ref{eq:Pot-r-w}), the $r^{p}$-potential $\mathrm{Pot}^{r^{p}}$ is linear in
the game. Moreover, the mapping $\mathrm{E}_{p}$ is linear in the game. Since
TUX Dirac games constitute a basis, (\ref{eq:dirac-rep}), it suffices to show
the claim for TUX Dirac games. For $N\subseteq\mathbf{U}$ and $\left(
T,\tau\right)  \in\mathcal{E}\left(  N\right)  ,$ $T\neq\emptyset,$ it follows
with the definition of the auxiliary game $v_{w}^{r}$ (\ref{eq:w-tilde-r})
that
\begin{equation}
v_{\delta_{T,\tau}}^{r^{p}}\overset{\text{(\ref{eq:w-tilde-r}%
),(\ref{eq:dirac-r-p-S})}}{=}\prod_{k=0}^{n-t-1}\frac{n-k}{n-t-k}\frac
{p_{N}\left(  \left\{  T\right\}  \cup\tau\right)  }{p_{T}\left(  \left\{
T\right\}  \right)  }\delta_{T}^{N}=\frac{n!}{\left(  n-t\right)  !t!}%
\frac{p_{N}\left(  \left\{  T\right\}  \cup\tau\right)  }{\frac{1}{t}}%
\delta_{T}^{N}, \label{eq:thm-r-p-1}%
\end{equation}
where we use \textbf{GEN} of $p$ and Corollary~\ref{cor:consequence-of-gen}%
~$\left(  i\right)  ,\ $i.e., $p_{T}\left(  \left\{  T\right\}  \right)
=1/t$, in the last equation. Now, we obtain%
\begin{align*}
\mathrm{Pot}^{r^{p}}\left(  \delta_{T,\tau}\right)   &  \overset
{\text{(\ref{eq:Pot-r-w})}}{=}\mathrm{Pot}\left(  v_{\delta_{T,\tau}}^{r^{p}%
}\right) \\
&  \overset{\text{(\ref{eq:thm-r-p-1})}}{=}\frac{n!t}{\left(  n-t\right)
!t!}p_{N}\left(  \left\{  T\right\}  \cup\tau\right)  \mathrm{Pot}\left(
\delta_{T}^{N}\right) \\
&  \overset{\text{(\ref{eq:pot-TU-dirac})}}{=}p_{N}\left(  \left\{  T\right\}
\cup\tau\right) \\
&  \overset{\text{(\ref{eq:def-Ep})}}{=}\mathrm{E}_{p}\left(  \delta_{T,\tau
}\right)  ,
\end{align*}
where we use the definition of the $r$-potential (\ref{eq:Pot-r-w}) in the
first equation, linearity of the potential for TU\ games (\ref{eq:thm-r-p-1})
in the second equation, the formula for the potential of Dirac TU\ games
(\ref{eq:pot-TU-dirac}) in the third equation, and the definition of
$\mathrm{E}_{p}$ (\ref{eq:def-Ep}) in the fourth equation. This concludes the
proof of existence.

\emph{Uniqueness\textit{\/}}:\textit{\emph{\/}} Let $r$ be a restriction
operator that satisfies \textbf{PI}, \textbf{PNG}, and (\textbf{\dag})
$\mathrm{Pot}^{r}=\mathrm{E}_{p}.$ We show $r=r^{p}$ by induction on $n.$

\emph{Induction basis}\textit{\emph{\/}}: For $w\in\mathbb{W}\left(  N\right)
,$ $N\subseteq\mathbf{U},$ $n=1,$ we have $w_{-i}^{r}=0=w_{-i}^{r^{p}}$ for
$i\in N$, where $\left\{  0\right\}  =\mathbb{W}\left(  \emptyset\right)  .$

\emph{Induction hypothesis (IH)}\textit{\emph{\/}}: Suppose we have
$w_{-i}^{r}=w_{-i}^{r^{p}}$ for all $w\in\mathbb{W}\left(  N\right)  ,$
$N\subseteq\mathbf{U}$ and $i\in N$ such that $n\leq m.$

\emph{Induction step}\textit{\emph{\/}}: Let $N\subseteq\mathbf{U}$ be such
that $n=m+1.$ For all $w\in\mathbb{W}\left(  N\right)  ,$ $i\in N,$ and
$\left(  T,\tau\right)  \in\mathcal{E}\left(  N\setminus\left\{  i\right\}
\right)  ,$ $T\neq\emptyset$, let $w_{i,T,\tau}\in\mathbb{W}\left(  N\right)
$ be given by%
\begin{equation}
w_{i,T,\tau}=\sum_{B\in\tau\cup\left\{  \emptyset\right\}  }w\left(
T,\tau_{+i\rightsquigarrow B}\right)  \delta_{T,\tau_{+i\rightsquigarrow B}}.
\label{eq:w-i-T-tau}%
\end{equation}
Note that in the game $w_{i,T,\tau}$ those embedded coalitions for $N$ that
determine $w_{-i}^{r}\left(  T,\tau\right)  $ according to the definition of a
restriction operator (\textbf{RES}) generate the same worth as in $w$, whereas
all other embedded coalition generate a zero worth. This is what drives the
results in the next two paragraphs---in the restricted game, only a few
embedded coalitions generate a worth that differs from their worth in the
restriction of the null game, which by \textbf{PNG} is the null game.

We have%
\begin{equation}
\left(  w_{i,T,\tau}\right)  _{-i}^{r}\overset{\text{(\ref{eq:w-i-T-tau}%
),\textbf{RES}}}{=}\xi\left(  w_{i,T,\tau},i\right)  \delta_{T,\tau}+\left(
\mathbf{0}^{N}\right)  _{-i}^{r}\overset{\text{\textbf{PNG}}}{=}\xi\left(
w_{i,T,\tau},i\right)  \delta_{T,\tau} \label{eq:w-i-T-tau-r}%
\end{equation}
for some $\xi\left(  w_{i,T,\tau},i\right)  \in\mathbb{R};$ and further using
the definition of TUX Dirac games (\ref{eq:dirac}), we have%
\begin{equation}
\left(  w_{i,T,\tau}\right)  _{-i}^{r}\left(  T,\tau\right)  \overset
{\text{(\ref{eq:dirac})}}{=}\xi\left(  w_{i,T,\tau},i\right)  .
\label{eq:w-i-T-tau-r(T,tau)}%
\end{equation}
For $k\in T,$ we have%
\begin{equation}
\left(  w_{i,T,\tau}\right)  _{-k}^{r}\overset{\text{(\ref{eq:w-i-T-tau}%
),\textbf{RES}}}{=}\left(  \mathbf{0}^{N}\right)  _{-k}^{r}\overset
{\text{\textbf{PNG}}}{=}\mathbf{0}^{N\setminus\left\{  k\right\}  }.
\label{eq:w-i-T-tau-r-k-0}%
\end{equation}

For $k\in\left(  N\setminus\left\{  i\right\}  \right)  \setminus T,$ we have%
\begin{align}
\left(  w_{i,T,\tau}\right)  _{-k}^{r}  &  \overset{\text{(\ref{eq:w-i-T-tau}%
),\textbf{RES}}}{=}\sum_{B\in\tau_{-k}\cup\left\{  \emptyset\right\}  }%
\xi\left(  w_{i,T,\tau},B,k\right)  \delta_{T,\left(  \tau_{_{-k}}\right)
_{+i\rightsquigarrow B}}+\left(  \mathbf{0}^{N}\right)  _{-k}^{r}\nonumber\\
&  \overset{\text{\textbf{PNG}}}{=}\sum_{B\in\tau_{-k}\cup\left\{
\emptyset\right\}  }\xi\left(  w_{i,T,\tau},B,k\right)  \delta_{T,\left(
\tau_{_{-k}}\right)  _{+i\rightsquigarrow B}} \label{eq:w-i-T-tau-r-k}%
\end{align}
for some $\xi\left(  w_{i,T,\tau},B,k\right)  \in\mathbb{R}.$ Hence, we obtain%
\begin{align}
\left(  \left(  w_{i,T,\tau}\right)  _{-i}^{r}\right)  _{-k}^{r}  &
\overset{\text{\emph{IH}}}{=}\left(  \left(  w_{i,T,\tau}\right)  _{-i}%
^{r}\right)  _{-k}^{r^{p}}\nonumber\\
&  \overset{\text{(\ref{eq:w-i-T-tau-r})}}{=}\left(  \xi\left(  w_{i,T,\tau
},i\right)  \delta_{T,\tau}\right)  _{-k}^{r^{p}}\nonumber\\
&  \overset{\text{(\ref{eq:r-p}),(\ref{eq:dirac-r-p})}}{=}\xi\left(
w_{i,T,\tau},i\right)  \frac{n-1}{n-1-t}\frac{p_{N\setminus\left\{  i\right\}
}\left(  \left\{  T\right\}  \cup\tau\right)  }{p_{N\setminus\left\{
i,k\right\}  }\left(  \left\{  T\right\}  \cup\tau_{-k}\right)  }%
\delta_{T,\tau_{-k}}, \label{eq:canext-1}%
\end{align}
where we use linearity following from the definition of the $r^{p}$
restriction operator (\ref{eq:r-p}) and the restriction of TUX Dirac games
(\ref{eq:dirac-r-p}) in the third equation. Analogously, we have
\begin{align}
&  \left(  \left(  w_{i,T,\tau}\right)  _{-k}^{r}\right)  _{-i}^{r}\nonumber\\
&  \overset{\text{\emph{IH}}}{=}\left(  \left(  w_{i,T,\tau}\right)  _{-k}%
^{r}\right)  _{-i}^{r^{p}}\nonumber\\
&  \overset{\text{(\ref{eq:w-i-T-tau-r-k})}}{=}\left(  \sum_{B\in\tau_{-k}%
\cup\left\{  \emptyset\right\}  }\xi\left(  w_{i,T,\tau},B,k\right)
\delta_{T,\left(  \tau_{_{-k}}\right)  _{+i\rightsquigarrow B}}\right)
_{-i}^{r^{p}}\nonumber\\
&  \overset{\text{(\ref{eq:r-p}),(\ref{eq:dirac-r-p})}}{=}\sum_{B\in\tau
_{-k}\cup\left\{  \emptyset\right\}  }\xi\left(  w_{i,T,\tau},B,k\right)
\frac{n}{n-t}\frac{p_{\left(  N\setminus\left\{  k\right\}  \right)  }\left(
\left\{  T\right\}  \cup\left(  \tau_{_{-k}}\right)  _{+i\rightsquigarrow
B}\right)  }{p_{N\setminus\left\{  i,k\right\}  }\left(  \left\{  T\right\}
\cup\left(  \left(  \tau_{_{-k}}\right)  _{+i\rightsquigarrow B}\right)
_{-i}\right)  }\delta_{T,\left(  \left(  \tau_{_{-k}}\right)
_{+i\rightsquigarrow B}\right)  _{-i}}\nonumber\\
&  =\left(  \sum_{B\in\tau_{-k}\cup\left\{  \emptyset\right\}  }\frac
{n-1}{n-1-t}\frac{p_{\left(  N\setminus\left\{  k\right\}  \right)  }\left(
\left\{  T\right\}  \cup\left(  \tau_{_{-k}}\right)  _{+i\rightsquigarrow
B}\right)  }{p_{N\setminus\left\{  i,k\right\}  }\left(  \left\{  T\right\}
\cup\tau_{_{-k}}\right)  }\xi\left(  w_{i,T,\tau},B,k\right)  \right)
\delta_{T,\tau_{-k}}. \label{eq:canext-2}%
\end{align}
Since $r$ satisfies \textbf{PI}, the left-hand sides of (\ref{eq:canext-1})
and (\ref{eq:canext-2}) coincide. We obtain%
\begin{align}
&  p_{N\setminus\left\{  i\right\}  }\left(  \left\{  T\right\}  \cup
\tau\right)  \xi\left(  w_{i,T,\tau},i\right) \nonumber\\
&  \qquad\overset{\text{(\ref{eq:canext-1}),(\ref{eq:canext-2})}}{=}\sum
_{B\in\tau_{-k}\cup\left\{  \emptyset\right\}  }p_{\left(  N\setminus\left\{
k\right\}  \right)  }\left(  \left\{  T\right\}  \cup\left(  \tau_{_{-k}%
}\right)  _{+i\rightsquigarrow B}\right)  \xi\left(  w_{i,T,\tau},B,k\right)
\label{eq:xi-i=xi-k}%
\end{align}
for all $k\in\left(  N\setminus\left\{  i\right\}  \right)  \setminus T.$

Further, we obtain%
\begin{align}
0  &  \overset{\text{(\ref{eq:w-i-T-tau})}}{=}w_{i,T,\tau}\left(
N,\emptyset\right) \nonumber\\
&  \overset{\text{(\ref{eq:r-pot2}),(\textbf{\dag})}}{=}\sum_{\ell\in
N}\left[  \mathrm{E}_{p}\left(  w_{i,T,\tau}\right)  -\mathrm{E}_{p}\left(
\left(  w_{i,T,\tau}\right)  _{-\ell}^{r}\right)  \right] \nonumber\\
&  \overset{\text{(\ref{eq:w-i-T-tau-r}),(\ref{eq:w-i-T-tau-r-k}%
),(\ref{eq:w-i-T-tau-r-k-0})}}{=}n\mathrm{E}_{p}\left(  w_{i,T,\tau}\right)
-\mathrm{E}_{p}\left(  \xi\left(  w_{i,T,\tau},i\right)  \delta_{T,\tau
}\right)  -\sum_{\ell\in T}\mathrm{E}_{p}\left(  \mathbf{0}^{N\setminus
\left\{  l\right\}  }\right) \nonumber\\
&  \qquad-\sum_{\ell\in\left(  N\setminus\left\{  i\right\}  \right)
\setminus T}\mathrm{E}_{p}\left(  \sum_{B\in\tau_{-\ell}\cup\left\{
\emptyset\right\}  }\xi\left(  w_{i,T,\tau},B,\ell\right)  \delta_{T,\left(
\tau_{-\ell}\right)  _{_{+i\rightsquigarrow B}}}\right) \nonumber\\
&  \overset{\text{(\ref{eq:def-Ep}),(\ref{eq:dirac})}}{=}n\mathrm{E}%
_{p}\left(  w_{i,T,\tau}\right)  -\xi\left(  w_{i,T,\tau},i\right)
p_{N\setminus\left\{  i\right\}  }\left(  \tau\cup\left\{  T\right\}  \right)
\nonumber\\
&  \qquad-\sum_{\ell\in\left(  N\setminus\left\{  i\right\}  \right)
\setminus T}\sum_{B\in\tau_{-\ell}\cup\left\{  \emptyset\right\}  }\xi\left(
w_{i,T,\tau},B,\ell\right)  p_{N\setminus\left\{  \ell\right\}  }\left(
\left(  \tau_{-\ell}\right)  _{_{+i\rightsquigarrow B}}\cup\left\{  T\right\}
\right) \nonumber\\
&  \overset{\text{(\ref{eq:xi-i=xi-k})}}{=}n\mathrm{E}_{p}\left(  w_{i,T,\tau
}\right)  -\left(  n-t\right)  p_{N\setminus\left\{  i\right\}  }\left(
\left\{  T\right\}  \cup\tau\right)  \xi\left(  w_{i,T,\tau},i\right)  ,
\label{eq:crucial}%
\end{align}
where in the second equation we use the formula for the potential
(\ref{eq:r-pot2}) and the fact that $\mathrm{Pot}^{r}=\mathrm{E}_{p}$; in the
the fourth equation refer to the definition of $\mathrm{E}_{p}$
(\ref{eq:def-Ep}) and the definition of TUX dirac games (\ref{eq:dirac}).
Finally, we obtain
\begin{align*}
w_{-i}^{r}\left(  T,\tau\right)   &  \overset{\text{\textbf{R}%
,(\ref{eq:w-i-T-tau})}}{=}\left(  w_{i,T,\tau}\right)  _{-i}^{r}\left(
T,\tau\right) \\
&  \overset{\text{(\ref{eq:w-i-T-tau-r(T,tau)})}}{=}\xi\left(  w_{i,T,\tau
},i\right) \\
&  \overset{\text{(\ref{eq:crucial})}}{=}\frac{n}{n-t}\frac{1}{p_{N\setminus
\left\{  i\right\}  }\left(  \left\{  T\right\}  \cup\tau\right)  }%
\mathrm{E}_{p}\left(  w_{i,T,\tau}\right) \\
&  \overset{\text{(\ref{eq:def-Ep}),(\ref{eq:w-i-T-tau}),(\ref{eq:dirac})}}%
{=}=\frac{n}{n-t}\sum_{B\in\tau\cup\left\{  \emptyset\right\}  }w\left(
T,\tau_{+i\rightsquigarrow B}\right)  \frac{p_{N}\left(  \left\{  T\right\}
\cup\tau_{+i\rightsquigarrow B}\right)  }{p_{N\setminus\left\{  i\right\}
}\left(  \left\{  T\right\}  \cup\tau\right)  }\\
&  \overset{\text{(\ref{eq:r-p})}}{=}w_{-i}^{r^{p}}\left(  T,\tau\right)  ,
\end{align*}
where in the fourth equation we refer to the definition of $\mathrm{E}_{p}$
(\ref{eq:def-Ep}) and the definition of TUX Dirac games (\ref{eq:dirac}); and
in the fifth equation we refer to the definition of the $r^{p}$-restriction
operator (\ref{eq:r-p}). This concludes the proof.

\subsection{Proof of Corollary~\ref{cor:sh-rp}}

$\left(  i\right)  $ Note that all mappings involved are linear. Since the
Dirac TUX games form a basis of the $\mathbb{W}\left(  N\right)  $ (see
(\ref{eq:dirac-rep})), it suffices to establish the formula for Dirac TUX
games. For all $N\subseteq\mathbf{U}$, $i\in N,$ and $\left(  T,\tau\right)
\in\mathcal{E}\left(  N\right)  ,$ $T\neq\emptyset,$ we obtain%
\begin{equation}
\mathrm{Sh}_{i}^{r^{p}}\left(  \delta_{T,\tau}\right)  \overset
{\text{(\ref{eq:r-Shapleyvalue})}}{=}\mathrm{Sh}_{i}\left(  v_{\delta_{T,\tau
}}^{r^{p}}\right)  \overset{\text{(\ref{eq:thm-r-p-1}),(\ref{eq:Sh})}}%
{=}\left\{
\begin{array}
[c]{ll}%
p_{N}\left(  \left\{  T\right\}  \cup\tau\right)  , & i\in T,\\
-\frac{t}{n-t}p_{N}\left(  \left\{  T\right\}  \cup\tau\right)  , & i\in
N\setminus T
\end{array}
\right.  , \label{eq:thm-r-p-2}%
\end{equation}
where in the first equation is due to the definition of the $r$-Shapley value
(\ref{eq:r-Shapleyvalue}); and the second equation is due to the relationship
$v_{\delta_{T,\tau}}^{r^{p}}=\frac{n!t}{\left(  n-t\right)  !t!}p_{N}\left(
\left\{  T\right\}  \cup\tau\right)  \delta_{T}^{N}$ described in
(\ref{eq:thm-r-p-1}), and the evaluation of the Shapley value for TU\ Dirac
games, $\mathrm{Sh}_{i}\left(  \delta_{T}^{N}\right)  =\frac{\left(
t-1\right)  !\left(  n-t\right)  !}{n!}$. This verifies the formula for Dirac
TUX games.

$\left(  ii\right)  $ Since both solutions are linear, it suffices to show the
claim for Dirac TU games. For all $N\subseteq\mathbf{U},$ $T\subseteq N,$
$T\neq\emptyset,$ and $i\in T$, we obtain%
\begin{align*}
\mathrm{Sh}_{i}^{r^{p}}\left(  \delta_{T}^{N}\right)  =\mathrm{Sh}_{i}^{r^{p}%
}\left(  \sum_{\tau\in\Pi\left(  N\setminus T\right)  }\delta_{T,\tau}%
^{N}\right)   &  \overset{\text{(\ref{eq:thm-r-p-2})}}{=}\sum_{\tau\in
\Pi\left(  N\setminus T\right)  }p_{N}\left(  \left\{  T\right\}  \cup
\tau\right) \\
&  \overset{\text{Prop~\ref{pro:p=/p-star}~(ii)}}{=}\dfrac{\left(  n-t\right)
!\left(  t-1\right)  !}{n!}\\
&  \overset{\text{(\ref{eq:Sh})}}{=}\mathrm{Sh}_{i}\left(  \delta_{T}%
^{N}\right)  ,
\end{align*}
where the second equation follows from Proposition~\ref{pro:p=/p-star}~(ii)
and the last equation follows from the definition of the\ Shapley value
(\ref{eq:Sh}). Analogously for $i\in N\setminus T.$

\subsection{Proof of Lemma~\ref{lem:rpstar}}

By (\ref{eq:p-star}), we have%
\[
\frac{1}{n-s}=\frac{n}{n-s}\frac{p_{N}^{\star}\left(  \left\{  S\right\}
\cup\pi_{+i\leadsto\emptyset}\right)  }{p_{N\setminus\left\{  i\right\}
}^{\star}\left(  \left\{  S\right\}  \cup\pi\right)  }\quad\text{and\quad
}\frac{b}{n-s}=\frac{n}{n-s}\frac{p_{N}^{\star}\left(  \left\{  S\right\}
\cup\pi_{+i\leadsto B}\right)  }{p_{N\setminus\left\{  i\right\}  }^{\star
}\left(  \left\{  S\right\}  \cup\pi\right)  }\quad\text{for }B\in\pi
\]
for all $N\subseteq\mathbf{U},$ $i\in N,$ and $\left(  S,\pi\right)
\in\mathcal{E}\left(  N\setminus\left\{  i\right\}  \right)  .$ The random
partition $p^{\star}$ is positive (\textbf{POS}) and generates the potential
for TU games (\textbf{GEN}). Hence, Theorem~\ref{thm:canext} applies. The
first claim now drops from (\ref{eq:r-star}) and (\ref{eq:r-p}).

Since both $\mathrm{MPW}$ and $\mathrm{Sh}^{r^{\star}}$ are linear, it
suffices to show the second claim for Dirac games. For all $N\subseteq
\mathbf{U},$ $T\subseteq N,$ and $i\in N,$ we have%
\begin{align*}
\mathrm{MPW}_{i}\left(  \delta_{T,\tau}\right)   &  \overset
{\text{(\ref{eq:Sh-star})}}{=}\mathrm{Sh}_{i}\left(  \bar{v}_{\delta_{T,\tau}%
}^{\star}\right) \\
&  \overset{\text{(\ref{eq:averagegame})}}{=}\mathrm{Sh}_{i}\left(
p_{N\setminus T}^{\star}\left(  \tau\right)  \delta_{T}^{N}\right) \\
&  \overset{\text{(\ref{eq:Sh})}}{=}\left\{
\begin{array}
[c]{ll}%
\frac{p_{N\setminus T}^{\star}\left(  \tau\right)  }{n\binom{n-1}{t-1}}, &
i\in T,\\
-\frac{t}{n-t}\frac{p_{N\setminus T}^{\star}\left(  \tau\right)  }%
{n\binom{n-1}{t-1}}, & i\in N\setminus T
\end{array}
\right. \\
&  \overset{\text{(\ref{eq:p-star})}}{=}\left\{
\begin{array}
[c]{ll}%
p_{N}^{\star}\left(  \left\{  T\right\}  \cup\tau\right)  , & i\in T,\\
-\frac{t}{n-t}p_{N}^{\star}\left(  \left\{  T\right\}  \cup\tau\right)  , &
i\in N\setminus T,
\end{array}
\right.
\end{align*}
where the first equation is due to the definition of $\mathrm{MPW}$
(\ref{eq:Sh-star}), the second equation is due to the definition of the
average game $\bar{v}^{\star}$ (\ref{eq:averagegame}), the third equation is
due to the definition of the Shapley value (\ref{eq:Sh}), and the fourth
equation is due to the definition of $p^{\star}$ (\ref{eq:p-star}). The claim
now follows from~(\ref{eq:thm-r-p-2}).

\subsection{Proof of Proposition~\ref{thm:rp+null=MPW}}

For all $N\subseteq\mathbf{U},$ $w\in\mathbb{W}\left(  N\right)  ,$ $i\in N,$
and $\left(  T,\tau\right)  \in\mathcal{E}\left(  N\setminus\left\{
i\right\}  \right)  ,$ $T\neq\emptyset,$ we have%
\begin{equation}
p_{N}^{\star}\left(  \left\{  T\cup\left\{  i\right\}  \right\}  \cup
\tau\right)  \overset{\text{(\ref{eq:p-star})}}{=}\frac{t}{n-t}\sum_{B\in
\tau\cup\left\{  \emptyset\right\}  }p_{N}^{\star}\left(  \left\{  T\right\}
\cup\tau_{+i\leadsto B}\right)  . \label{eq:SLE-p-star}%
\end{equation}
Hence, we obtain%
\[
\mathrm{Sh}_{i}^{r^{p^{\star}}}\left(  w\right)  =\sum_{\left(  T,\tau\right)
\in\mathcal{E}\left(  N\setminus\left\{  i\right\}  \right)  }\left(  \frac
{t}{n-t}\sum_{B\in\tau\cup\left\{  \emptyset\right\}  }p_{N}^{\star}\left(
\left\{  T\right\}  \cup\tau_{+i\leadsto B}\right)  \left[  w\left(
T\cup\left\{  i\right\}  ,\tau\right)  -w\left(  T,\tau_{+i\leadsto B}\right)
\right]  \right)
\]
for all $N\subseteq\mathbf{U},$ $w\in\mathbb{W}\left(  N\right)  ,$ and $i\in
N.$ This already entails that $\mathrm{Sh}^{r^{p^{\star}}}$ satisfies
\textbf{NP}.

Let now $p$ be a random partition. If $n\leq1,$ then trivially $p_{N}%
=p_{N}^{\star}.$ Let now $N\subseteq\mathbf{U}$ be such that $n>1.$ For
$\alpha\in\mathbb{R},$ $i\in N$, $\pi\in\Pi\left(  N\setminus\left\{
i\right\}  \right)  ,$ and $B\in\Pi,$ let $w_{i,\pi,B}^{\alpha}\in
\mathbb{W}\left(  N\right)  $ be given by%
\[
w_{i,\pi,B}^{\alpha}\equiv\alpha\delta_{B\cup\left\{  i\right\}  ,\pi
\setminus\left\{  B\right\}  }+\sum_{C\in\left(  \pi\setminus\left\{
B\right\}  \right)  \cup\left\{  \emptyset\right\}  }\alpha\delta_{B,\left(
\pi\setminus\left\{  B\right\}  \right)  _{+i\leadsto C}}.
\]
By construction, player $i$ is a null player in $w_{i,\pi,B}^{\alpha}$ for any
$\alpha\in\mathbb{R}.$ By \textbf{NP}, we therefore have $\mathrm{Sh}_{i}%
^{p}\left(  w_{i,\pi,B}^{1}\right)  =0.$ By (\ref{eq:Sh-rp}), we have%
\[
\mathrm{Sh}_{i}^{p}\left(  w_{i,\pi,B}^{\alpha}\right)  =\alpha\left(
p_{N}\left(  \pi_{+i\leadsto B}\right)  -\frac{b}{n-b}\sum_{C\in\left(
\pi\setminus\left\{  \pi\left(  i\right)  \right\}  \right)  \cup\left\{
\emptyset\right\}  }p_{N}\left(  \pi_{+i\leadsto C}\right)  \right)  .
\]
Hence, \textbf{NP} implies%
\begin{equation}
p_{N}\left(  \pi_{+i\leadsto B}\right)  -\frac{b}{n-b}\sum_{C\in\left(
\pi\setminus\left\{  B\right\}  \right)  \cup\left\{  \emptyset\right\}
}p_{N}\left(  \pi_{+i\leadsto C}\right)  =0. \label{eq:SLE-1}%
\end{equation}
By (\ref{eq:SLE-p-star}), the random partition $p^{\star}$ satisfies
(\ref{eq:SLE-1}).

Let $\beta\in\mathbb{R}$ be defined by $p_{N}\left(  \left[  N\right]
\right)  =\beta p_{N}^{\star}\left(  \left[  N\right]  \right)  =\frac{\beta
}{n!}.$ We show $p_{N}\left(  \pi\right)  =\beta\cdot p_{N}^{\star}\left(
\pi\right)  $ for all $\pi\in\Pi\left(  N\right)  $ by induction on
$\left\vert \pi\right\vert .$ Since both $p_{N}$ and $p_{N}^{\star}$ are
probability distributions on $\Pi\left(  N\right)  ,$ this implies
$p_{N}=p_{N}^{\star}$.

\emph{Induction basis}\textit{\emph{\/}}: By construction, the claim holds for
$\left\vert \pi\right\vert =n,$ that is, $\pi=\left[  N\right]  .$

\emph{Induction hypothesis (IH)}\textit{\emph{\/}}: Let the claim hold for all
$\pi\in$ $\Pi\left(  N\right)  $ with $\left\vert \pi\right\vert \geq t>1.$

\emph{Induction step}\textit{\emph{\/}}: Fix $i\in N.$ Let now $\pi\in$
$\Pi\left(  N\setminus\left\{  i\right\}  \right)  $ be such that $\left\vert
\pi\right\vert =t-1.$ By (\ref{eq:SLE-1}), we have%
\begin{equation}
p_{N}\left(  \pi_{+i\leadsto B}\right)  -\frac{b}{n-b}\sum_{C\in\pi
\setminus\left\{  B\right\}  }p_{N}\left(  \pi_{+i\leadsto C}\right)
=\frac{b}{n-b}p_{N}\left(  \pi_{+i\leadsto\emptyset}\right)  \overset
{\text{\emph{IH}}}{=}\frac{b\cdot\beta}{n-b}p_{N}^{\star}\left(
\pi_{+i\leadsto\emptyset}\right)  \label{eq:SLE-2}%
\end{equation}
for all $B\in\pi.$ That is, we have a system of $\left\vert \pi\right\vert $
linear equations indexed by $B\in\pi$ with $\left\vert \pi\right\vert $
unknowns, $p_{N}\left(  \pi_{+i\leadsto C}\right)  ,$ indexed by $C\in\pi.$
This system be written as $Ax=y.$ The column vector $x\equiv\left(
p_{N}\left(  \pi_{+i\leadsto C}\right)  \right)  _{C\in\pi}$ represents the
$\left\vert \pi\right\vert $ unknowns. The column vector $y\equiv\left(
\frac{b\cdot\beta}{n-b}p_{N}^{\star}\left(  \pi_{+i\leadsto\emptyset}\right)
\right)  _{B\in\pi}$ represents the right-hand side of (\ref{eq:SLE-2}). The
$\left\vert \pi\right\vert \times\left\vert \pi\right\vert $ matrix
$A\equiv\left(  a_{B,C}\right)  _{B,C\in\pi}$ given by%
\[
a_{B,C}\equiv\left\{
\begin{array}
[c]{ll}%
1, & C=B,\\
-\frac{b}{n-b}, & C\neq B
\end{array}
\right.
\]
represents the coefficients of (\ref{eq:SLE-2}), where the left index
indicates the equation and the right index indicates the unknown. Note that we
cover all $\pi\in\Pi\left(  N\right)  $ with $\left\vert \pi\right\vert =t-1.$

By (\ref{eq:SLE-p-star}), the column vector $x^{\star}\equiv\left(  \beta\cdot
p_{N}^{\star}\left(  \pi_{+i\leadsto C}\right)  \right)  _{C\in\pi}$ is a
solution of (\ref{eq:SLE-2}). Remains to show that $x^{\star}$ is the unique
solution. This is the case if and only if $A$ is non-singular. Consider the
$\left\vert \pi\right\vert \times\left\vert \pi\right\vert $ matrix $D=\left(
d_{B,C}\right)  _{B,C\in\pi}$ given by%
\[
d_{B,C}\equiv\left\{
\begin{array}
[c]{ll}%
p_{N}^{\star}\left(  \pi_{+i\leadsto B}\right)  , & C=B,\\
0, & C\neq B.
\end{array}
\right.
\]
This matrix is non-singular. Hence, the matrix $A$ is non-singular if and only
if the product $AD=\left(  e_{B,C}\right)  _{B,C\in\pi}$ given by%
\begin{equation}
e_{B,C}\equiv\left\{
\begin{array}
[c]{ll}%
p_{N}^{\star}\left(  \pi_{+i\leadsto B}\right)  , & C=B,\\
-\frac{b}{n-b}p_{N}^{\star}\left(  \pi_{+i\leadsto C}\right)  , & C\neq B
\end{array}
\right.  \label{eq:AD}%
\end{equation}
is non-singular. For any $B\in\pi,$ we have%
\[
\left\vert e_{B,B}\right\vert -\sum_{C\in\pi\setminus\left\{  B\right\}
}\left\vert e_{B,C}\right\vert \overset{\text{(\ref{eq:AD})}}{=}p_{N}^{\star
}\left(  \pi_{+i\leadsto B}\right)  -\frac{b}{n-b}\sum_{C\in\pi\setminus
\left\{  B\right\}  }p_{N}^{\star}\left(  \pi_{+i\leadsto C}\right)
\overset{\text{(\ref{eq:SLE-1})}}{=}\frac{b}{n-b}p_{N}^{\star}\left(
\pi_{+i\leadsto\emptyset}\right)  \overset{\text{(\ref{eq:p-star})}}{>}0,
\]
where the last inequality is follows from the definition of $p^{\star}$.
Hence, the matrix $AD$ has a strictly dominant diagonal. It is well-known that
real square matrices with a strictly dominant diagonal are non-singular
\citep[see for example][Theorem~1]{taussky1949}%
, which concludes the proof.

\subsection{Proof of Theorem~\ref{thm:MPW}}

The proof of this theorem heavily relies on arguments used to show the
preceding results in this paper. In order to avoid massive repetitions, we
just make the sketch provided at the end of Subsection~\ref{sec:MPW} as
precise as possible given its (necessary) vagueness. An attentive reader
should be able to easily fill in all the details.

\emph{Existence}\textit{\emph{\/}}: By Theorems~\ref{thm:Pot-p-star}
and~\ref{thm:canext} and Lemma~\ref{lem:rpstar}, the random partition
$p^{\star}$ and the restriction operator $r^{\star}$ satisfy the properties of
the theorem.

\emph{Uniqueness}\textit{\emph{\/}}: Let $p$ be a random partition satisfying
\textbf{GEN} (generation of the potential for TU games) and let $r$ be a
path-independent (\textbf{PI}) restriction operator that preserves null games
(\textbf{PNG}). Moreover, let the $r$-potential coincide with the expected
accumulated worth of $p$, $\mathrm{Pot}^{r}=\mathrm{E}_{p}$, and let the
corresponding $r$-Shapley value $\mathrm{Sh}^{r}$ satisfy the null player
property (\textbf{NP}) or marginality (\textbf{M}). We show $p_{N}%
=p_{N}^{\star}$ for all $N\subseteq\mathbf{U}$ by induction on $n.$ The rest
of the claim then follows from Theorem~\ref{thm:canext},
Corollary~\ref{cor:sh-rp}, and Lemma~\ref{lem:rpstar}.

\emph{Induction basis}\textit{\emph{\/}}: For $n\leq3$, $p_{N}=p_{N}^{\star}$
follows from \textbf{GEN} of $p$ and Corollary~\ref{cor:consequence-of-gen}.

\emph{Induction hypothesis (IH)}\textit{\emph{\/}}: Suppose $p_{N}%
=p_{N}^{\star}$ for all $N\subseteq\mathbf{U}$ such that $n\leq a.$

\emph{Induction step}\textit{\emph{\/}}: Let $N\subseteq\mathbf{U}$ be such
that $n\leq a+1.$ Note that the enumeration below follows the enumeration in
the sketch provided at the end of Subsection~\ref{sec:MPW}.

\begin{enumerate}
\item[(i)] By the induction hypothesis (\emph{IH}\textit{\emph{\/}}) and the
definition of $p^{\star}$ (\ref{eq:p-star}), we have $p_{T}\left(  \pi\right)
=p_{T}^{\star}\left(  \pi\right)  >0$ for all $T\subseteq N$ with $t\leq n-1$
and $\pi\in\Pi\left(  T\right)  $. Hence, the restricted games $w_{-i}^{r^{p}%
}$ as in (\ref{eq:r-p}) are well-defined for all $T\subseteq N,$
$w\in\mathbb{W}\left(  T\right)  ,$ and $i\in T$.

\item[(ii)] Running the arguments in the proof of Theorem~\ref{thm:canext} for
$N$ instead of $\mathbf{U}$ shows $w^{r}=w^{r^{p}}$ for all $T\subseteq N,$
$w\in\mathbb{W}\left(  T\right)  .$

\item[(iii)] Running the arguments in the proof of Corollary~\ref{cor:sh-rp}
for $N$ instead of $\mathbf{U}$ shows $\mathrm{Sh}^{r}\left(  w\right)
=\mathrm{Sh}^{r^{p}}\left(  w\right)  =\mathrm{Sh}^{p}\left(  w\right)  $ for
all $T\subseteq N,$ $w\in\mathbb{W}\left(  T\right)  .$

\item[(iv)] Running the arguments in the proof of
Theorem~\ref{thm:rp+null=MPW} or Corollary~\ref{cor:rp+mo=MPW} for $N$ instead
of $\mathbf{U}$ shows in particular $p_{N}=p_{N}^{\star}.$
\end{enumerate}

\subsection{Proof of Proposition~\ref{pro:CI+GEN=/p-star}}

\emph{Existence}\textit{\emph{\/}}: By Theorem~\ref{thm:Pot-p-star}, the
random partition $p^{\star}$ satisfies \textbf{GEN}. The Ewens distributions
defined in Footnote~\ref{fn:ewens}, in particular $p^{\star},$ are known to
satisfy \textbf{CI}.

\emph{Uniqueness}\textit{\emph{\/}}: Let the random partition $p$ for
$\mathbf{U}$ satisfy \textbf{CI} and \textbf{GEN}. We show $p_{N}=p_{N}%
^{\star}$ for all $N\subseteq\mathbf{U}$ by induction on$~n.$

\emph{Induction basis}\textit{\emph{\/}}: For $n=1$, the claim is immediate.

\emph{Induction hypothesis (IH)}\textit{\emph{\/}}: Suppose $p_{N}%
=p_{N}^{\star}$ for all $N\subseteq\mathbf{U}$ such that $n\leq\bar{n}.$

\emph{Induction step}\textit{\emph{\/}}: Let $N\subseteq\mathbf{U}$ be such
that $n=\bar{n}+1.$ We have%
\[
p_{N}\left(  \left\{  N\right\}  \right)  \overset{\text{\textbf{GEN}}}%
{=}\mathrm{Pot}\left(  u_{N}^{N}\right)  \overset{\text{(\ref{eq:Pot-p-star}%
)}}{=}p_{N}^{\star}\left(  \left\{  N\right\}  \right)  ,
\]
where the second equation is due to the fact that $p^{\star}$ generates the
potential (\ref{eq:Pot-p-star}). Consider now $\pi\in\Pi\left(  N\right)  ,$
$\pi\neq\left\{  N\right\}  $ and $B\in\pi.$ We obtain%
\begin{align*}
p_{N}\left(  \pi\right)   &  \overset{\text{\textbf{CI}}}{=}p_{N\setminus
B}\left(  \pi\setminus\left\{  B\right\}  \right)  \sum_{\tau\in\Pi\left(
N\right)  :B\in\tau}p_{N}\left(  \tau\right) \\
&  \overset{\text{\textbf{GEN}}}{=}p_{N\setminus B}\left(  \pi\setminus
\left\{  B\right\}  \right)  \mathrm{Pot}\left(  \delta_{B}^{N}\right) \\
&  \overset{\text{\emph{IH}}}{=}p_{N\setminus B}^{\star}\left(  \pi
\setminus\left\{  B\right\}  \right)  \mathrm{Pot}\left(  \delta_{B}%
^{N}\right) \\
&  \overset{\text{(\ref{eq:Pot-p-star})}}{=}p_{N\setminus B}^{\star}\left(
\pi\setminus\left\{  B\right\}  \right)  \sum_{\tau\in\Pi\left(  N\right)
:B\in\tau}p_{N}^{\star}\left(  \tau\right) \\
&  \overset{\text{\textbf{CI}}}{=}p_{N}^{\star}\left(  \pi\right)  ,
\end{align*}
which concludes the proof.

\subsection{Proof of Proposition~\ref{pro:w-bar-star-strong-null}}

We prepare the proof by a lemma.

\begin{lemma}
\label{lem:commute}For all $w\in\mathbb{W}\left(  N\right)  \mathbf{,}$
$N\subseteq\mathbf{U}$ and $i\in N$, we have $\bar{v}_{w_{-i}^{r^{\star}}%
}^{\star}=\left(  \bar{v}_{w}^{\star}\right)  _{-i}$.
\end{lemma}

\noindent\textbf{Proof.}$\;$The claim follows from the linearity of the
involved operators, from
\[
\bar{v}_{\left(  \delta_{T,\tau}\right)  _{-i}^{r^{\star}}}^{\star}%
\overset{\text{(\ref{eq:dirac-r-p})}}{=}\bar{v}_{\delta_{T,\tau_{-i}}}^{\star
}\overset{\text{(\ref{eq:averagegame})}}{=}\delta_{T}^{N\setminus\left\{
i\right\}  }\overset{\text{(\ref{eq:averagegame})}}{=}\left(  \bar{v}%
_{\delta_{T,\tau}}^{\star}\right)  _{-i}%
\]
for all $N\subseteq\mathbf{U},$ $T\subseteq N,$ $T\neq\emptyset,$ $\left(
T,\tau\right)  \in\mathcal{E}\left(  N\right)  ,$ and $i\in N\setminus
T,\ $and%
\[
\bar{v}_{\left(  \delta_{T,\tau}\right)  _{-i}^{r^{\star}}}^{\star}%
\overset{\text{(\ref{eq:dirac-r-p})}}{=}\bar{v}_{\mathbf{0}^{N\setminus
\left\{  i\right\}  }}^{\star}\overset{\text{(\ref{eq:averagegame})}}%
{=}\mathbf{0}^{N\setminus\left\{  i\right\}  }\overset
{\text{(\ref{eq:averagegame})}}{=}\left(  \bar{v}_{\delta_{T,\tau}}^{\star
}\right)  _{-i}%
\]
for $i\in T$, where we first use the restriction formula for TUX Dirac games
(\ref{eq:dirac-r-p}) and then definition of the average
game(\ref{eq:averagegame}).~\hspace*{\fill}$\square$

We can now prove the proposition.

\noindent\textbf{Proof of Proposition~\ref{pro:w-bar-star-strong-null}.}%
$\;$For all $w\in\mathbb{W}\left(  N\right)  ,$ $N\subseteq\mathbf{U}$ and
$i,j\in N,$ $i\neq j$ such that $i$ is a strong null player in $w,$ we have%
\[
\mathrm{MPW}_{i}\left(  w_{-j}^{r^{\star}}\right)  \overset
{\text{(\ref{eq:Sh-star})}}{=}\mathrm{Sh}_{i}\left(  \bar{v}_{w_{-j}%
^{r^{\star}}}^{\star}\right)  \overset{\text{Lemma~\ref{lem:commute}}}%
{=}\mathrm{Sh}_{i}\left(  \left(  \bar{v}_{w}^{\star}\right)  _{-j}\right)
=0,
\]
where the first equation just follows from the definition of $\mathrm{MPW}$
and the third equation follows from the facts that the strong null player $i$
becomes a null player in the average game $\bar{v}_{w}^{\star}$
\citep[p.~346]{MSPCWe2007}%
, that null players remain null players in a subgame of a TU game, and that
null players obtain a zero Shapley payoff.~\hspace*{\fill}$\square$

\bibliographystyle{elsart-harv}
\bibliography{iddmpw_AC}

\newcommand{\noopsort}[1]{}
\begin{thebibliography}{42}
\expandafter\ifx\csname natexlab\endcsname\relax\def\natexlab#1{#1}\fi
\expandafter\ifx\csname url\endcsname\relax
  \def\url#1{\texttt{#1}}\fi
\expandafter\ifx\csname urlprefix\endcsname\relax\def\urlprefix{URL }\fi

\bibitem[{Albizuri et~al.(2005)Albizuri, Arin, and Rubio}]{AlArRu2005}
Albizuri, M.~J., Arin, J., Rubio, J., 2005. An axiom system for a value for
  games in partition function form. International Game Theory Review 7, 63--73.

\bibitem[{Aldous(1985)}]{aldous1985}
Aldous, D., 1985. Exchangeability and related topics. In: Hennequin, P.~L.
  (Ed.), Ecole d'été de probabilités de Saint-Flour, XIII---1983. Vol. 1117 of
  Lecture Notes in Mathematics. Springer, Berlin, Ch.~1, pp. 1--198.

\bibitem[{Bolger(1989)}]{bolger1989}
Bolger, E.~M., 1989. A set of axioms for a value for partition function games.
  International Journal of Game Theory 18, 37--44.

\bibitem[{Br{\"u}gemann et~al.(2018)Br{\"u}gemann, Gautier, and
  Menzio}]{BruGauMen2018}
Br{\"u}gemann, B., Gautier, P., Menzio, G., 2018. Intra firm bargaining and
  {Shapley} values. The Review of Economic Studies 86~(2), 564--592.

\bibitem[{Calvo and Santos(1997)}]{CalSan1997}
Calvo, E., Santos, J.~C., 1997. Potentials in cooperative {TU} games.
  Mathematical Social Sciences 34, 175--190.

\bibitem[{Casajus(2014)}]{casajus-potential}
Casajus, A., 2014. Potential, value, and random partitions. Economics Letters
  126~(2), 164--166.

\bibitem[{Casajus and Huettner(2018)}]{CasHue-deva}
Casajus, A., Huettner, F., 2018. Decomposition of solutions and the {Shapley}
  value. Games and Economic Behavior 13~(3), 1--23.

\bibitem[{Castro et~al.(2009)Castro, Gómez, and Tejada}]{CaGoTe2009}
Castro, J., Gómez, D., Tejada, J., 2009. Polynomial calculation of the
  {Shapley} value based on sampling. Computers and Operations Research 36~(5),
  1726--1730.

\bibitem[{{\noopsort{Clippel}}{de Clippel} and Serrano(2008)}]{CliSer2008}
{\noopsort{Clippel}}{de Clippel}, G., Serrano, R., 2008. Marginal contributions
  and externalities in the value. Econometrica 76~(6), 1413--1436.

\bibitem[{Crane(2016)}]{crane2016}
Crane, H., 2016. The ubiquitous {Ewens} sampling formula. Statistical Science
  31~(1), 1--19.

\bibitem[{Dutta et~al.(2010)Dutta, Ehlers, and Kar}]{DuEhKa2010}
Dutta, B., Ehlers, L., Kar, A., 2010. Externalities, potential, value and
  consistency. Journal of Economic Theory 145~(6), 2380--2411.

\bibitem[{Ewens(1972)}]{ewens1972}
Ewens, W., 1972. The sampling theory of selectively neutral alleles.
  Theoretical Population Biology 3, 87--112.

\bibitem[{Grabisch and Funaki(2012)}]{GraFun2012}
Grabisch, M., Funaki, Y., 2012. A coalition formation value for games in
  partition function form. European Journal of Operational Research 221~(1),
  175--185.

\bibitem[{Gul(1989)}]{gul1989}
Gul, F., 1989. Bargaining foundations of {Shapley} value. Econometrica 57~(1),
  81.

\bibitem[{Hart and {Mas-Colell}(1989)}]{HarMas1989}
Hart, S., {Mas-Colell}, A., 1989. Potential, value, and consistency.
  Econometrica 57~(3), 589--614.

\bibitem[{Hart and Mas-Colell(1996)}]{HarMas1996}
Hart, S., Mas-Colell, A., 1996. Bargaining and value. Econometrica 64~(2),
  357--380.

\bibitem[{Kamijo and Kongo(2012)}]{KamKon2012}
Kamijo, Y., Kongo, T., 2012. Whose deletion does not affect your payoff? {The}
  difference between the {Shapley} value, the egalitarian value, the solidarity
  value, and the {Banzhaf} value. European Journal of Operational Research
  216~(3), 638--646.

\bibitem[{Karlin and McGregor(1972)}]{KarMcG1972}
Karlin, S., McGregor, J., 1972. Addendum to a paper of {W.\ Ewens}. Theoretical
  Population Biology 3~(1), 113--116.

\bibitem[{Kingman(1978)}]{kingman1978}
Kingman, J. F.~C., 1978. Random partitions in population genetics. Proceeedings
  of the Royal Society of London. A. Mathematical and Physical Sciences
  361~(1704).

\bibitem[{Krishna and Serrano(1995)}]{KriSer1995}
Krishna, V., Serrano, R., 1995. Perfect equilibria of a model of n-person
  noncooperative bargaining. International Journal of Game Theory 24, 259--272.

\bibitem[{Kóczy(2018)}]{koszy2018}
Kóczy, L.~Á., 2018. Partition Function Form Games. Springer International
  Publishing.

\bibitem[{Lundberg et~al.(2020)Lundberg, Erion, Chen, DeGrave, Prutkin, Nair,
  Katz, Himmelfarb, Bansal, and Lee}]{Lundbergetal2020}
Lundberg, S.~M., Erion, G., Chen, H., DeGrave, A., Prutkin, J.~M., Nair, B.,
  Katz, R., Himmelfarb, J., Bansal, N., Lee, S.-I., 2020. From local
  explanations to global understanding with explainable {AI} for trees. Nature
  Machine Intelligence 2~(1), 56--67.

\bibitem[{Macho-Stadler et~al.(2007)Macho-Stadler, P{\'e}rez-Castrillo, and
  Wettstein}]{MSPCWe2007}
Macho-Stadler, I., P{\'e}rez-Castrillo, D., Wettstein, D., 2007. Sharing the
  surplus: An extension of the {Shapley} value for environments with
  externalities. Journal of Economic Theory 135~(1), 339--356.

\bibitem[{McQuillin(2009)}]{mcquillin2009}
McQuillin, B., 2009. The extended and generalized {Shapley} value:
  {Simultaneous} consideration of coalitional externalities and coalitional
  structure. Journal of Economic Theory 144, 696--721.

\bibitem[{McQuillin and Sugden(2016)}]{McQSug2016}
McQuillin, B., Sugden, R., 2016. Backward induction foundations of the
  {Shapley} value. Econometrica 84~(6), 2265--2280.

\bibitem[{Michalak et~al.(2013)Michalak, Aadithya, Szczepanski, Ravindran, and
  Jennings}]{MiAaSzRaJe2013}
Michalak, T.~P., Aadithya, K.~V., Szczepanski, P.~L., Ravindran, B., Jennings,
  N.~R., 2013. Efficient computation of the {Shapley} value for game-theoretic
  network centrality. Journal of Artificial Intelligence Research 46, 607--650.

\bibitem[{Myerson(1977)}]{myerson1977pffg}
Myerson, R.~B., 1977. Values of games in partition function form. International
  Journal of Game Theory 6~(1), 23--31.

\bibitem[{Myerson(1980)}]{myerson1980}
Myerson, R.~B., 1980. Conference structures and fair allocation rules.
  International Journal of Game Theory 9, 169--182.

\bibitem[{Ortmann(1998)}]{ortmann1998}
Ortmann, K.~M., 1998. Conservation of energy in value theory. Mathematical
  Methods of Operations Research 47, 423--449.

\bibitem[{{Pham Do} and Norde(2007)}]{PhaNor2007}
{Pham Do}, K.~H., Norde, H., 2007. The {Shapley} value for partition function
  form games. International Game Theory Review 9, 353--360.

\bibitem[{Pitman(2006)}]{pitman2006}
Pitman, J., 2006. Combinatorial Stochastic Processes. Springer.

\bibitem[{Pérez-Castrillo and Wettstein(2001)}]{PerWet2001}
Pérez-Castrillo, D., Wettstein, D., 2001. Bidding for the surplus: A
  non-cooperative approach to the {Shapley} value. Journal of Economic Theory
  100~(2), 274--294.

\bibitem[{S{\'{a}}nchez-P{\'{e}}rez(2016)}]{sanchezperez2017}
S{\'{a}}nchez-P{\'{e}}rez, J., 2016. A decomposition for the space of games
  with externalities. International Journal of Game Theory 46~(1), 205--233.

\bibitem[{Shapley(1953)}]{shapley1953}
Shapley, L.~S., 1953. A value for $n$-person games. In: Kuhn, H., Tucker, A.
  (Eds.), Contributions to the Theory of Games. Vol.~II. Princeton University
  Press, Princeton, pp. 307--317.

\bibitem[{Skibski and Michalak(2019)}]{SkiMic2020}
Skibski, O., Michalak, T., 2019. Fair division in the presence of
  externalities. International Journal of Game Theory 49~(1), 147--172.

\bibitem[{Skibski et~al.(2018)Skibski, Michalak, and Woolbridge}]{SkMiWo2018}
Skibski, O., Michalak, T.~P., Woolbridge, M., 2018. The stochastic {Shapley}
  value for coalitional games with externalities. Games and Economic Bahavior
  108, 65--80.

\bibitem[{Sobolev(1975)}]{sobolev1975}
Sobolev, A.~I., 1975. The characterization of optimality principles in
  cooperative games by functional equations. In: Vorobev, N.~N. (Ed.),
  Mathematical Methods in the Social Sciences. Vol.~6. Academy of Sciences of
  the Lithuanian SSR, Vilnius, pp. 95--151, in Russian.

\bibitem[{Stole and Zwiebel(1996)}]{StoZwi1996}
Stole, L.~A., Zwiebel, J., 1996. Intra-firm bargaining under non-binding
  contracts. The Review of Economic Studies 63~(3), 375.

\bibitem[{Taussky(1949)}]{taussky1949}
Taussky, O., 1949. A recurring theorem on determinants. The American
  Mathematical Monthly 56~(10), 672--676.

\bibitem[{Thrall and Lucas(1963)}]{ThrLuc1963}
Thrall, R.~M., Lucas, W.~F., 1963. $n$-person games in partition function form.
  Naval Research Logistic Quarterly 10, 281--293.

\bibitem[{Yang et~al.(2019)Yang, Sun, Hou, and Xu}]{YaSuHoXu2019}
Yang, G., Sun, H., Hou, D., Xu, G., 2019. Games in sequencing situations with
  externalities. European Journal of Operational Research 278~(2), 699--708.

\bibitem[{Young(1985)}]{young1985}
Young, H.~P., 1985. Monotonic solutions of cooperative games. International
  Journal of Game Theory 14, 65--72.

\end{thebibliography}







\end{document}